\documentclass[11pt,english]{article}
\usepackage{lmodern}

\usepackage[T1]{fontenc}
\usepackage[latin9]{inputenc}
\usepackage{geometry}
\usepackage{xcolor}
\usepackage{verbatim}
\usepackage{float}
\usepackage{textcomp}
\usepackage{amsmath}
\usepackage{amsthm}
\usepackage{amssymb}
\usepackage{graphicx}
\usepackage{setspace}
\usepackage[authoryear]{natbib}
\makeatletter

\newcommand{\lyxdot}{.}

\theoremstyle{plain}
\newtheorem{thm}{\protect\theoremname}
\theoremstyle{definition}
\newtheorem{defn}{\protect\definitionname}
\theoremstyle{plain}
\newtheorem{cor}{\protect\corollaryname}

\setcitestyle{round}
\usepackage{lscape}
\usepackage{longtable}

\usepackage{graphicx}
\usepackage{tabularx}

\usepackage{array,booktabs,ragged2e}
\newcolumntype{C}[1]{>{\centering\arraybackslash}p{#1}}
\newcolumntype{J}[1]{>{\justify\arraybackslash}p{#1}}
\newcolumntype{R}[1]{>{\RaggedLeft\arraybackslash}p{#1}}
\newcolumntype{Q}[1]{>{\columncolor{Gray}\RaggedLeft\arraybackslash}p{#1}}
\newcolumntype{L}[1]{>{\RaggedRight\arraybackslash}p{#1}}
\newcolumntype{G}{@{\extracolsep{0.5cm}}l@{\extracolsep{0pt}}}%
\usepackage{multirow}
\usepackage{centernot}

\@ifundefined{showcaptionsetup}{}{%
 \PassOptionsToPackage{caption=false}{subfig}}
\usepackage{subfig}
\AtBeginDocument{

}

\makeatother

\usepackage{babel}
\providecommand{\corollaryname}{Corollary}
\providecommand{\definitionname}{Definition}
\providecommand{\theoremname}{Theorem}

\begin{document}
\title{A New Method for Generating Random Correlation Matrices\thanks{We thank seminar participants at Duke University, Penn State University,
Emory University, and the University of Notre Dame for helpful comments
and Megan Mccoy for proofreading the first draft.}\emph{\normalsize{}\medskip{}
}}
\author{\textbf{Ilya Archakov}$^{a}$\textbf{ and Peter Reinhard Hansen}$^{b,c}$\textbf{
and Yiyao Luo}$^{b}$\textbf{}\thanks{Address: University of North Carolina, Department of Economics, 107
Gardner Hall Chapel Hill, NC 27599-3305}\bigskip{}
\\
{\normalsize{}$^{a}$}\emph{\normalsize{}University of Vienna\smallskip{}
}\\
{\normalsize{}$^{b}$}\emph{\normalsize{}University of North Carolina\smallskip{}
}\\
{\normalsize{}$^{c}$}\emph{\normalsize{}Copenhagen Business School\medskip{}
}}
\date{\emph{\normalsize{}\today}}
\maketitle
\begin{abstract}
We propose a new method for generating random correlation matrices
that makes it simple to control both location and dispersion. The
method is based on a vector parameterization, $\gamma=g(C)$, which
maps any distribution on $\mathbb{R}^{n(n-1)/2}$ to a distribution
on the space of non-singular $n\times n$ correlation matrices. Correlation
matrices with certain properties, such as being well-conditioned,
having block structures, and having strictly positive elements, are
simple to generate. We compare the new method with existing methods.

\bigskip{}
\end{abstract}
\textit{\small{}Keywords:}{\small{} Random Correlation Matrix, Fisher
Transformation, Covariance Modeling.}{\small\par}

\noindent \textit{\small{}JEL Classification:}{\small{} C10; C15;
C58 \newpage}{\small\par}

\section{Introduction}

The correlation matrix plays a central role in many multivariate models.
Random correlation matrices are commonly used in Bayesian analysis
to specify priors, in multivariate probit models, and to investigate
the properties of estimators and hypotheses tests. Generating random
$n\times n$ correlation matrices can become onerous if the correlation
matrix is required to have certain features, such as non-negative
correlations or a block structure. Several distinct methods were proposed
in the literature to serve different needs, see \citet{Pourahmadi2011}
for a review. In this paper, we propose a novel method for generating
random correlation matrices, which is well-suited for a wide range
of objectives. The new method can, in principle, be used to generate
random correlation matrices with any distribution on the set on non-singular
correlation matrices. Positive definite correlation matrices are guaranteed,
and it is simple to control both the location and dispersion of the
correlation matrix. It is also simple to generate random correlation
matrices in the vicinity of a particular correlation matrix. We characterize
a way to generate a broad class of homogeneous distributions. This
refers to the case where the distribution is invariant to reordering
of the variables, and one implication of this invariance is that the
marginal distributions for the individual correlations are identical.
We also show how a heterogeneous random correlation matrix can be
generated, which refers to the the case where some correlation coefficients
are more disburse than other coefficients. An inequality makes it
straight forward to bound the smallest eigenvalue of the random correlation
matrix. The new method also makes it simple to generate random correlation
matrices with some special structures, such as block structures or
with strictly positive coefficients.

The rest of this paper is organized as follows. We introduce the new
method for generating random correlation matrices in Section 2 and
discuss several features and structures that can be generated with
the new method in Section 3. In Section 4, we review some existing
methods for generating random correlation matrices and discuss their
properties. We summarize in Section 5, present proofs in Appendix
A, and some auxiliary results in Appendix B.

\section{Random Correlation Matrices: A New Method}

The proposed method for generating random correlation matrices is
based on the following vector parameterization of non-singular correlation
matrices, 
\begin{equation}
\gamma=g(C):=\mathrm{vecl}(\log C),\label{eq:MappingCtoGamma}
\end{equation}
where the operator $\mathrm{vecl}(\cdot)$ vectorizes the lower off-diagonal
elements and $\log C$ is the matrix logarithm of $C$.\footnote{The matrix logarithm for a non-singular correlation matrix with eigendecomposition,
$C=Q\Lambda Q^{\prime}$, is given by $\log C=Q\log\Lambda Q^{\prime},$
where $\log\Lambda=\mathrm{diag}(\log\lambda_{1},\ldots,\log\lambda_{n})$.} The mapping, $g$, is a one-to-one correspondence between the set
of $n\times n$ non-singular correlation matrices, denoted $\mathcal{C}_{n\times n}$,
and $\mathbb{R}^{d}$, where $d=n(n-1)/2$, see \citet{ArchakovHansen:Correlation}.
So, any vector, $\gamma\in\mathbb{R}^{d}$, corresponds to a unique
correlation matrix $C(\gamma)\equiv g^{-1}(\gamma)$, and vice versa.

The new method for generating a random correlation matrix is simple:
it only requires computing $C(\gamma)$ from a random vector, $\gamma\in\mathbb{R}^{d}$.
The mapping, $\gamma\mapsto C(\gamma)$, will induce a distribution
on $\mathcal{C}_{n\times n}$ from any distribution on $\mathbb{R}^{d}$.
For instance, the density, $f_{\gamma}(\gamma)$, on $\mathbb{R}^{d}$,
will translate to the density
\begin{equation}
f_{C}(C)=f_{\gamma}(g(C))|\psi(C)|,\qquad\text{on }\mathcal{C}_{n\times n},\label{eq:density_transform}
\end{equation}
where $\psi(C)$ is the determinant of $\mathrm{d}\gamma/\mathrm{d}\varrho$
and $\varrho=\mathrm{vecl}C$ is the vector with the correlation coefficients
in $C$. An algorithm for computing $C(\gamma)$ and the determinant,
$\psi(C)$, is given in \citet{ArchakovHansen:Correlation}. A simple
example, for the case $n=2$, is the logistic density, $f_{\gamma}(\gamma)=2e^{-2\gamma}/(1+e^{-2\gamma})^{2}$,
which translates to a random $2\times2$ correlation matrix where
the correlation coefficient is uniformly distributed on $[-1,1]$.
This is a special case of Theorem \ref{thm:Logistic2Beta}, which
is presented in Section \ref{subsec:Equicorrelation-Matrices}.%

\subsection{Correlation Coefficients with Identical Marginal Distributions}

Some existing methods for generating random correlation matrices are
carefully crafted to generate correlation coefficients with identical
marginal distributions. \citet{Joe:2006} derived a method that yields
Beta distributed correlation coefficients on $[-1,1]$, and \citet{PourahmadiWang:2015}
arrived at the same result using a different approach. The new method
makes it possible to generate identically distributed coefficients
with a wide range of distributions beyond Beta distributions. For
instance, the correlations coefficients, $C_{ij}$, are identically
distributed whenever $\gamma_{i}$, $i=1,\ldots,d$, are independent
and identically distributed. Identically distributed correlations
can also be obtained with a common component and index-specific components
in the elements of $\gamma$.
\begin{thm}[Permutation invariance]
\label{thm:Permutations}Let $\gamma=\mathrm{vecl}(G)$, where $G_{ij}=h(\zeta,\xi_{i},\xi_{j},\varepsilon_{ij})$,
$1\leq j<i\leq n$, for some $h:\mathbb{R}^{4}\curvearrowright\text{\ensuremath{\mathbb{R}}}$.
If the three sets of variables, $\zeta$, $(\xi_{1},\ldots,\xi_{n})$,
and $\{\varepsilon_{ij}\}_{1\leq j<i\leq n}$, are mutually independent,
with $\varepsilon_{ij}$ independent and identically distributed,
and $\xi_{1},\ldots,\xi_{n}$ independent and identically distributed,
then $C(\gamma)$ and $\tilde{C}=PC(\gamma)P^{\prime}$ are identically
distributed on $\mathcal{C}_{n\times n}$ for any permutation matrix,
$P\in\mathbb{R}^{n\times n}$. 
\end{thm}
\begin{figure}[H]
\begin{centering}
\subfloat[$\gamma\sim N_{3}(0,I)$]{\centering{}\includegraphics[width=0.23\textwidth]{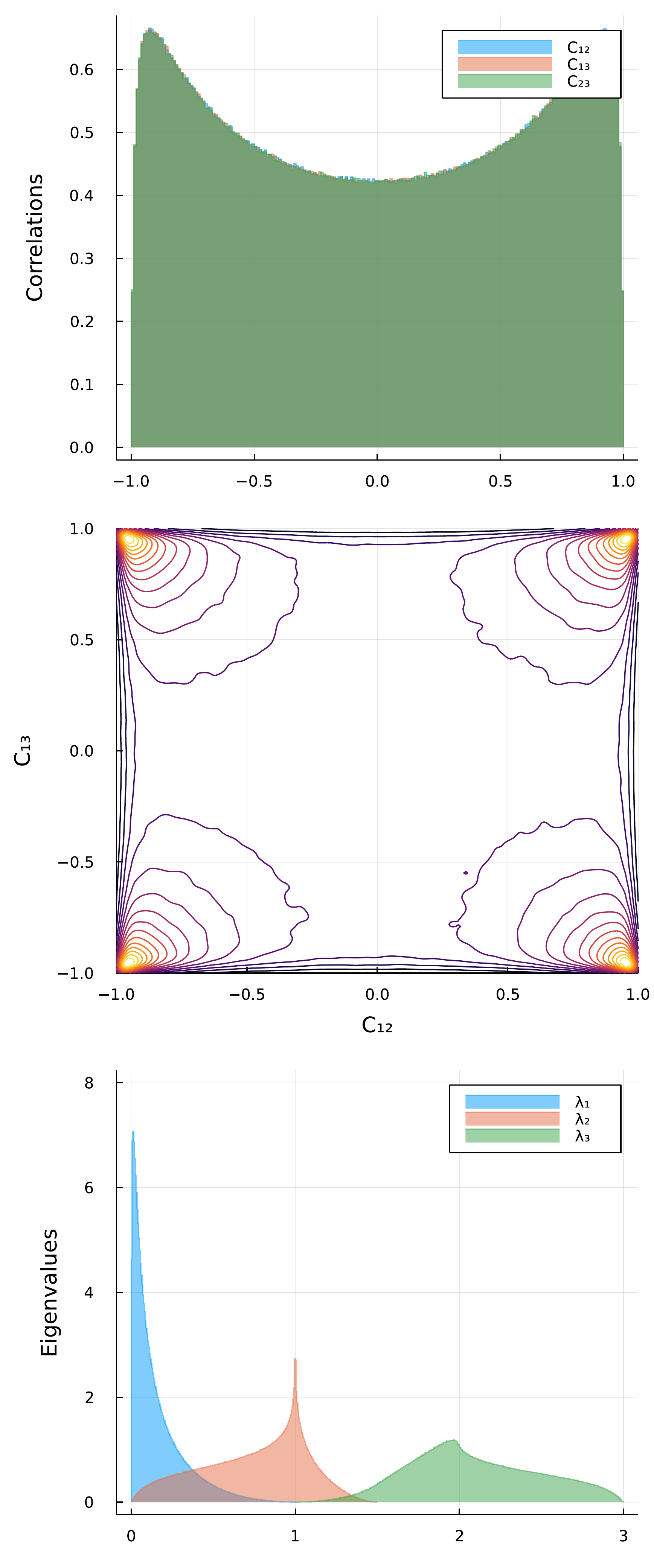}}\subfloat[$\gamma\sim N_{3}(0,\tfrac{1}{4}I)$]{\centering{}\includegraphics[width=0.23\textwidth]{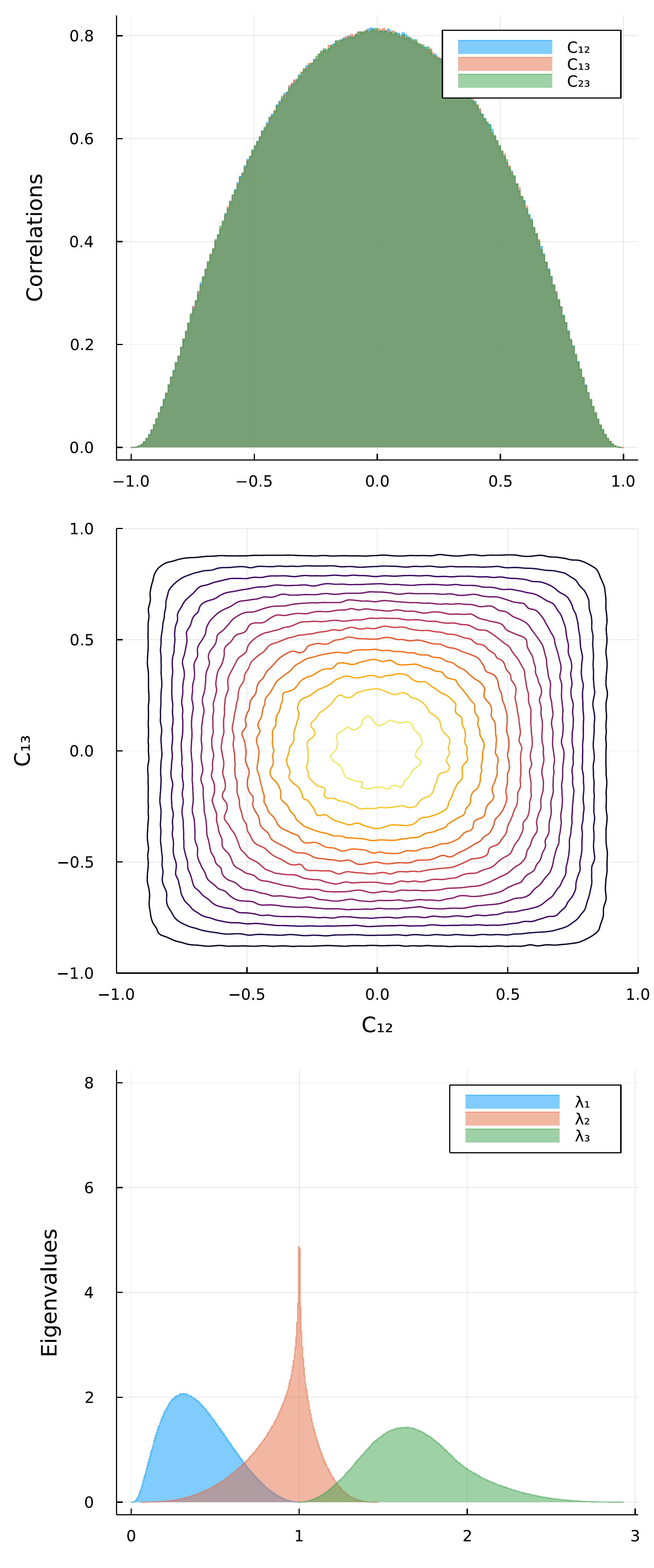}}\subfloat[$\gamma\sim N_{3}(0,\tfrac{1}{16}I)$]{\centering{}\includegraphics[width=0.23\textwidth]{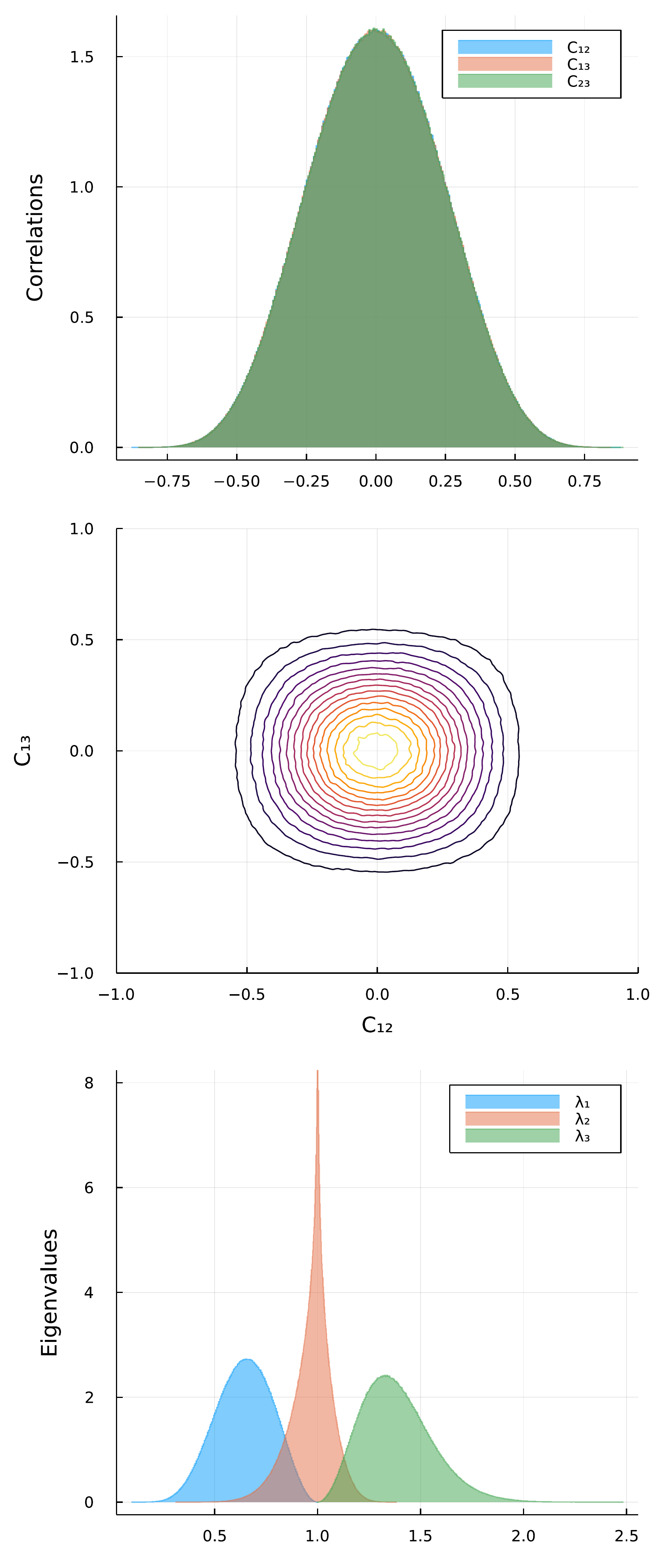}}\subfloat[$\gamma\sim N_{3}(0,\tfrac{1}{64}I)$]{\begin{centering}
\includegraphics[width=0.23\textwidth]{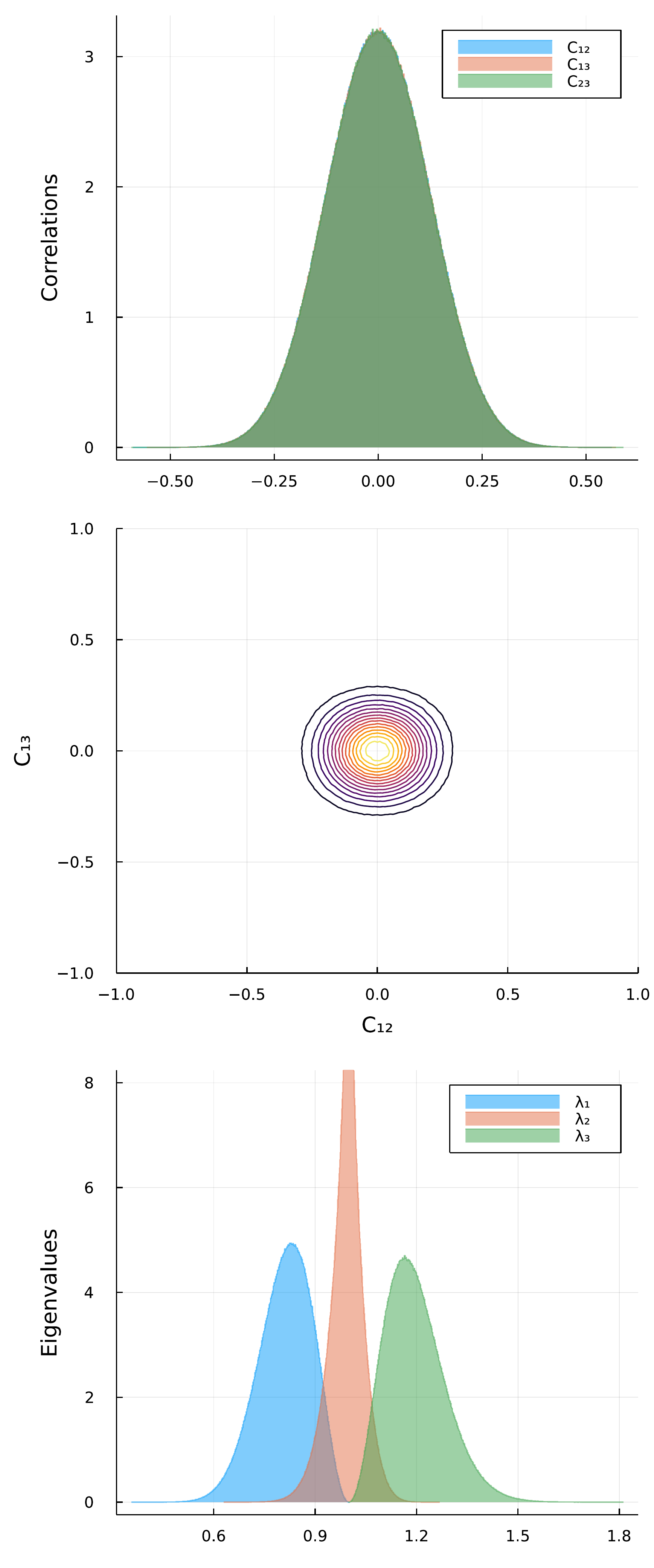}
\par\end{centering}
}
\par\end{centering}
\begin{centering}
\subfloat[$\gamma\sim N_{3}(\gamma_{0},I)$]{\centering{}\includegraphics[width=0.23\textwidth]{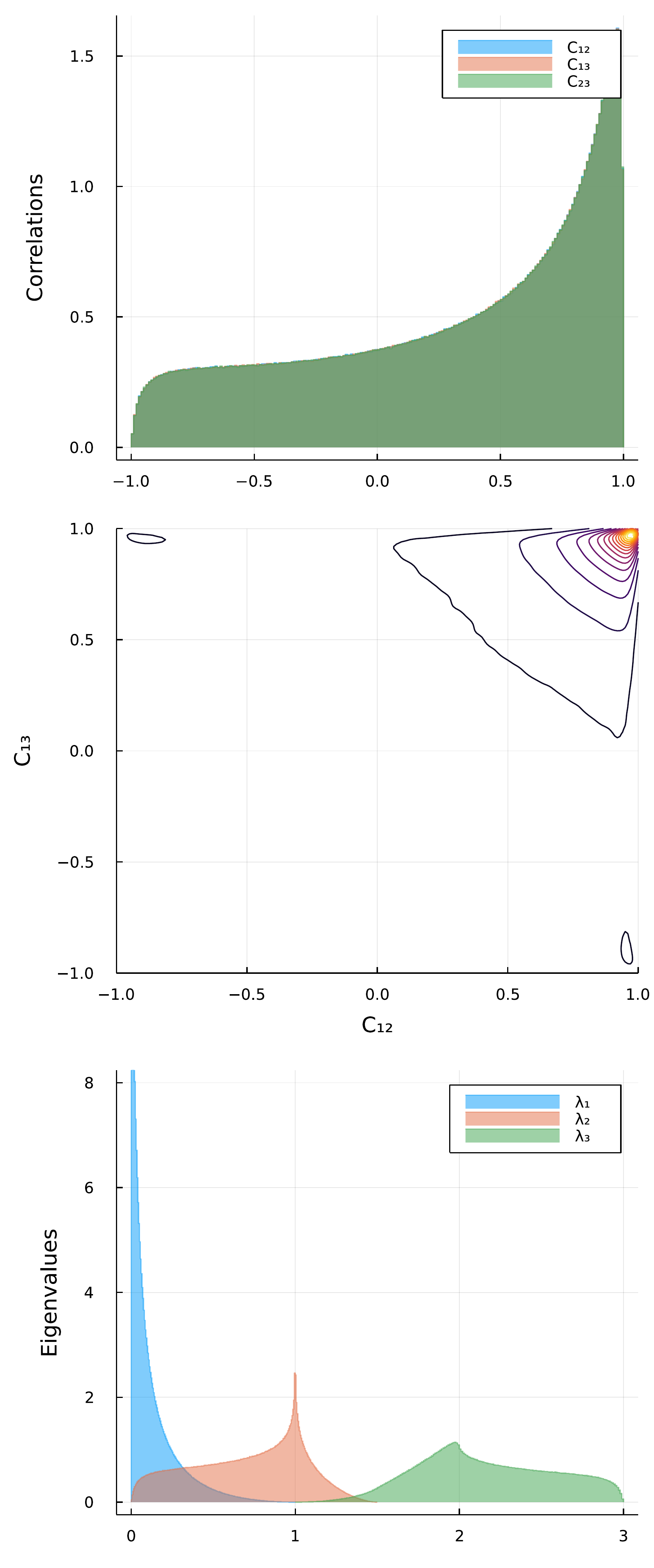}}\subfloat[$\gamma\sim N_{3}(\gamma_{0},\tfrac{1}{4}I)$]{\centering{}\includegraphics[width=0.23\textwidth]{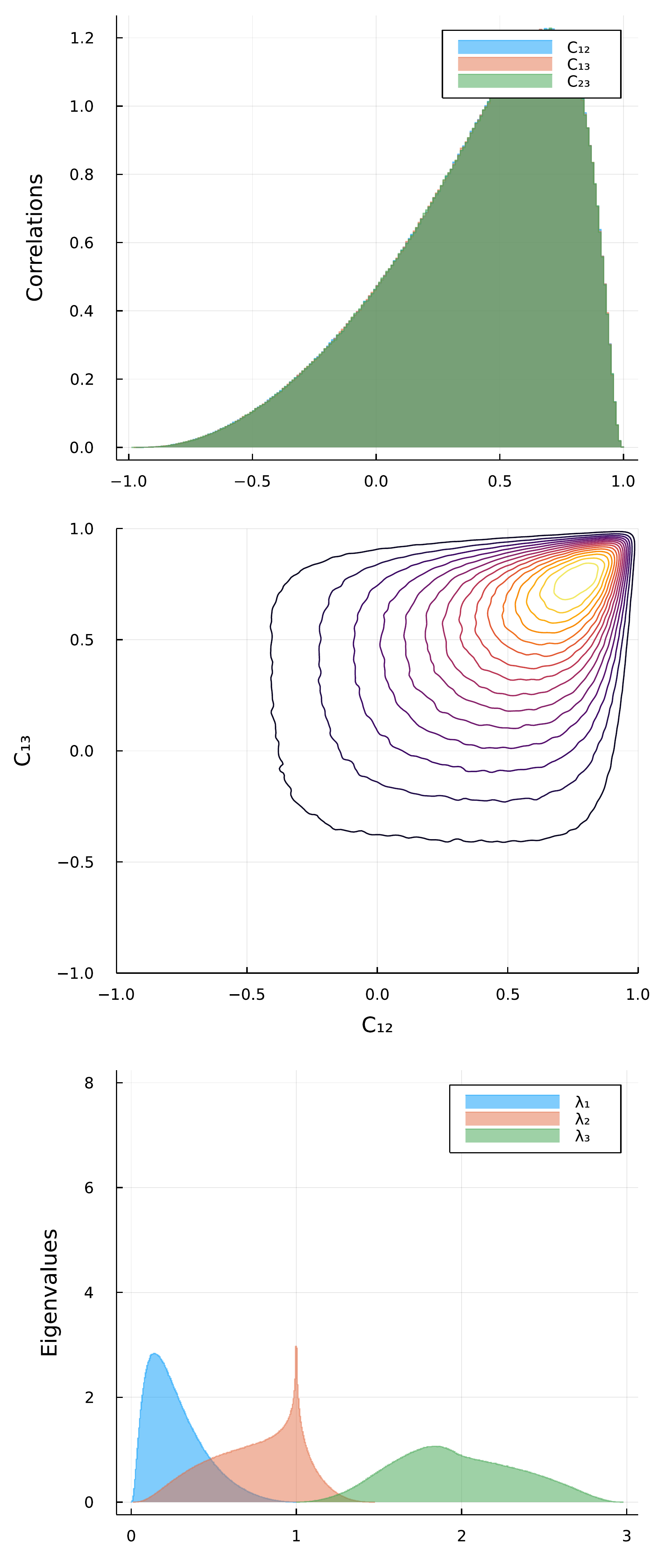}}\subfloat[$\gamma\sim N_{3}(\gamma_{0},\tfrac{1}{16}I)$]{\centering{}\includegraphics[width=0.23\textwidth]{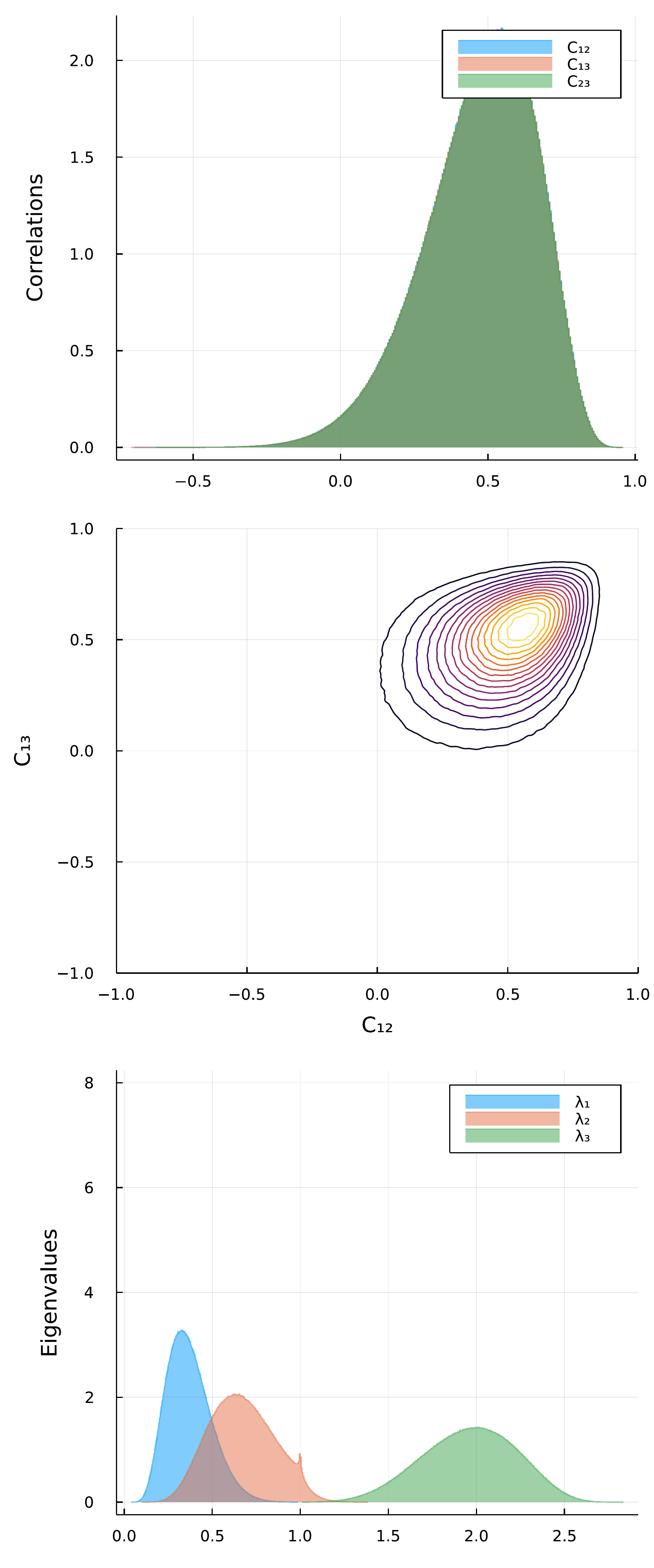}}\subfloat[$\gamma\sim N_{3}(\gamma_{0},\tfrac{1}{64}I)$]{\begin{centering}
\includegraphics[width=0.23\textwidth]{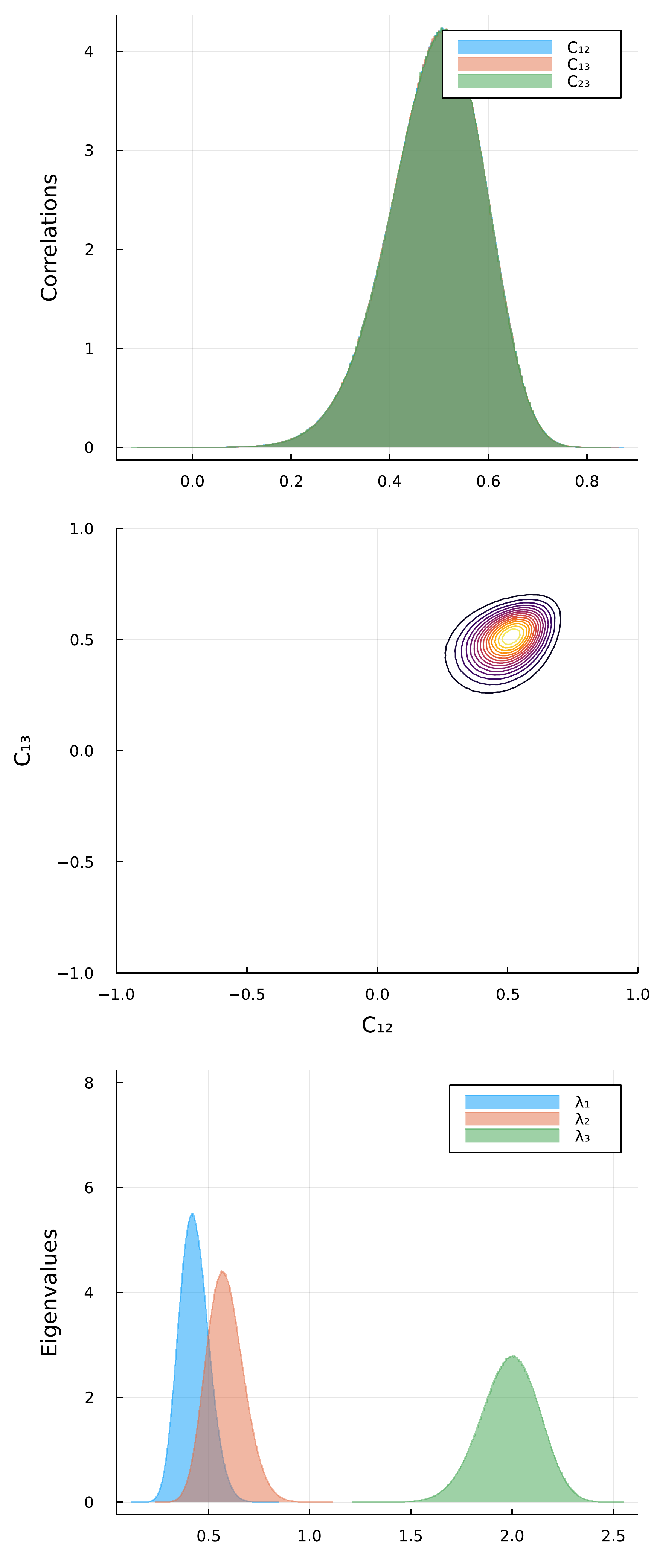}
\par\end{centering}
}
\par\end{centering}
\caption{Properties of random $3\times3$ correlation matrices generated from
$\gamma\sim N_{3}(\gamma_{0},\omega^{2}I)$ with $\gamma_{0}=(\mu,\mu,\mu)^{\prime}$.
Upper panels have $\mu=0$ and lower panels have $\mu=\tfrac{1}{3}\log4$.
Panels from left to right have $\omega^{2}=1,\tfrac{1}{4},\tfrac{1}{16}$,
and $\tfrac{1}{64}$ respectively. Each panel displays: the marginal
distribution of $C_{ij}$; contour plot for the bivariate distribution
of $(C_{12},C_{13})$; and the densities of ordered eigenvalues of
$C$.\label{fig:dim3_density-gamma}}
\end{figure}

An immediate implication of Theorem \ref{thm:Permutations} is that
all of the marginal distributions of the correlations, $C_{ij}$,
$i\neq j$ are identical under the stated assumptions. More generally,
the vector of correlations in the upper left $m\times m$ principal
submatrix, $\varrho=\mathrm{vecl}([C(\gamma)]_{i,j=1,\ldots,m})\in\mathbb{R}^{m(m-1)/2}$,
$m\leq n$, has the same distribution as the vector of correlations
corresponding to any other principal submatrix, $\tilde{\varrho}=\mathrm{vecl}([C(\gamma)]_{i,j\in\mathcal{I}})$,
for some $\mathcal{I}=\{i_{1},\ldots,i_{m}\}\subset\{1,\ldots,n\}$.
Under the conditions of Theorem \ref{thm:Permutations}, the pairs,
$(C_{12},C_{23})$, $(C_{12},C_{13})$, and $(C_{13},C_{23})$, have
the same bivariate distribution, but their bivariate distribution
need not be identical to that of $(C_{12},C_{34})$, because this
pair does not share a common index. 

The simplest case to consider in Theorem \ref{thm:Permutations} is
$\zeta=\xi_{1}=\cdots=\xi_{n}=0$, such that the element of $\gamma$
are independent and identically distributed. We illustrate the new
method for generating random correlation matrices by using this design
with independent and Gaussian distributed elements of $\gamma$. Some
features of the resulting random correlation matrices are shown in
Figure \ref{fig:dim3_density-gamma} for the case where $n=3$. Panels
(a)-(d) correspond to the case where $\gamma_{i}\sim iidN(0,\omega^{2})$,
$i=1,2,3$, such that the random correlation matrices are located
about $C=I_{3}$. Panels (e)-(h) are based on $\gamma_{i}\sim iidN(\tfrac{1}{3}\log4,\omega^{2})$.
This leads to random correlation matrices in the vicinity of 
\[
C(\gamma^{\ast})=\left[\begin{array}{ccc}
1 & 0.5 & 0.5\\
0.5 & 1 & 0.5\\
0.5 & 0.5 & 1
\end{array}\right],\qquad\text{where}\quad\gamma^{\ast}=\tfrac{\log4}{3}\left[\begin{array}{c}
1\\
1\\
1
\end{array}\right].
\]
In each panel of Figure \ref{fig:dim3_density-gamma}, we display
(from top to bottom) the marginal distributions for the correlation
coefficients, contour plots for bivariate distributions, and the densities
for the three eigenvalues. The panels in Figure \ref{fig:dim3_density-gamma}
correspond to the cases where $\omega^{2}=1$, $\tfrac{1}{4}$, $\tfrac{1}{16}$,
and $\tfrac{1}{64}$, respectively. From Theorem \ref{thm:Permutations}
we know that the marginal distributions are identical when the elements
of $\gamma$ are independent, and this can be seen from the simulated
densities for $C_{12}$, $C_{13}$, and $C_{23}$, that are indistinguishable
in all cases. The contour plots are for the bivariate distribution
of $(C_{12},C_{13})$, which are identical to the distributions for
any of pair of correlation coefficients as a consequence of Theorem
\ref{thm:Permutations}. 

When the variance of the elements of $\gamma$ is relatively large,
$\omega^{2}=1$, then $C(\gamma)$ tends to produce near-singular
correlation matrices. This is evident from the distribution of the
smallest eigenvalue in Panels (a) and (e), and it can also be seen
from the contour plots where the mass in concentrated near the corners
of the support for $(C_{12},C_{13})$. As the variance of $\gamma_{i}$
becomes smaller, so does the variance of the resulting correlation
coefficients. In Panels (a)-(d), the random correlation matrices become
more concentrated about $C(0)=I_{3}$ and in Panels (e)-(h) the random
correlations are more concentrated about $\frac{1}{2}$ as $\omega\rightarrow0$.

\subsection{Random Perturbation of Target Correlation Matrix}

The new method makes it easy to generate random correlation matrices
in the vicinity of a particular correlation matrix. Let $\gamma_{0}=g(C_{0})$
be the vector that corresponds to $C_{0}$ and generate random correlation
matrices using $C(\gamma_{0}+\epsilon)$, where $\epsilon$ is a random
vector centered about the zero-vector. The dispersion of the random
correlation matrices about $C_{0}$ is controlled by the dispersion
of $\varepsilon$. We will make use of this property below.

It is important to note that the random correlation matrices are unlikely
to have $\mathbb{E}(C)=C_{0}$, because the mapping $C(\gamma)$ is
non-linear. However, the discrepancy will be small if the variance
of $\varepsilon$ is small.

\subsection{\label{subsec:Heterogenous-Marginal-Distributi}Heterogenous Marginal
Distributions}

\begin{figure}[h]
\begin{centering}
\includegraphics[width=1\textwidth]{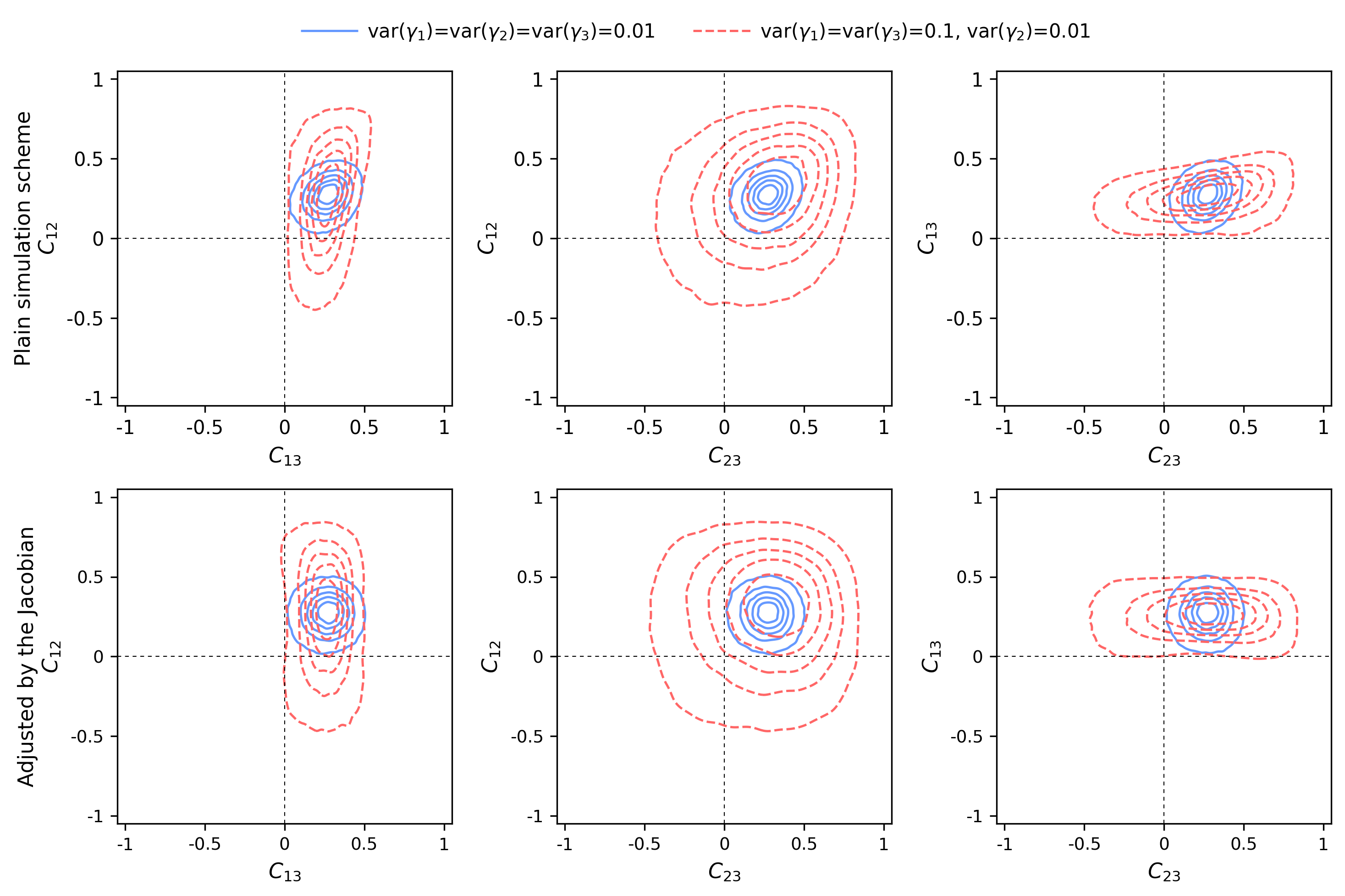}
\par\end{centering}
\caption{Contour plots for the bivariate distributions for $(C_{12},C_{13})$,
$(C_{12},C_{23})$, and $(C_{13},C_{23})$, where $C(\gamma)$ is
generated from $\gamma\sim N_{3}(\gamma_{0},\Sigma)$, with $\gamma_{0}=\tfrac{1}{4}(1,1,1)^{\prime}$.
Upper panels are with $\Sigma=\Lambda_{\omega}=\text{diag}(\tfrac{\omega}{100},\tfrac{1}{100},\tfrac{\omega}{100})$
and lower panels have $\Sigma=J_{0}^{-1}\Lambda_{\omega}J_{0}^{-1}$
where $J_{0}=\left.\mathrm{d}\varrho/\mathrm{d}\gamma\right|_{\gamma=\gamma_{0}}$
is the Jacobian. Solid blue contour lines are for the homogeneous
design with $\omega=1$ and red dashed contour lines are for the heterogeneous
case $\omega=10$. \label{fig:heter_gamma}}
\end{figure}

In some applications it can be desirable to generate random correlation
matrices where the dispersion of the correlation coefficients is heterogeneous.
This situation will arise in a Bayesian context if there is stronger
prior knowledge about some correlation coefficients than other correlations.
The new method can accommodate this situation by using different variances
for different elements in $\gamma$. The mapping in (\ref{eq:MappingCtoGamma})
is such that its Jacobian, $J_{0}=\mathrm{d}\varrho/\mathrm{d}\gamma\Bigl|_{\gamma=\gamma_{0}}$,
is approximately a diagonal matrix.\footnote{For examples, see \citet[figures S.6 and S.7]{ArchakovHansen:CorrAppendix},
who present the Jacobian matrices for a Toeplitz correlation matrix
and an empirical correlation matrix for daily industry portfolio returns.} Its diagonal elements are all positive and have similar magnitudes,
whereas the off-diagonal elements tend to be close to zero. So, increased
variance in a particular element of $\gamma$ will primarily induce
increased dispersion of the corresponding elements of $\varrho$.
For instance, increasing the variance of $\gamma_{1}=[\log C]_{1,2}$
will primarily increase the variance in $\varrho_{1}=C_{1,2}$. There
will also be an impact on other correlation coefficients for two reasons.
First, the Jacobian only captures a local linear approximation of
the mapping, $\gamma\mapsto\varrho=\mathrm{vecl}C(\gamma)$, and second,
$J_{0}$ is not perfectly diagonal. This is illustrated in the upper
panels of Figure \ref{fig:heter_gamma}. Random correlation matrices
were obtained with $\gamma\sim N_{3}(\gamma_{0},\Omega)$, where $\gamma_{0}=\Bigl(\frac{1}{4},\frac{1}{4},\frac{1}{4}\Bigl)^{\prime}$
and $\Omega=\text{diag}(\tfrac{\omega}{100},\tfrac{1}{100},\tfrac{\omega}{100})$.
The resulting contour plots for $(C_{12},C_{13})$, $(C_{12},C_{23})$,
and $(C_{13},C_{23})$ are show in the upper panels of Figure \ref{fig:heter_gamma},
where blue solid contour lines correspond to the homogeneous dispersion
$(\omega=1)$ and red dashed contour lines represent the heterogeneous
case, $\omega=10$, where $\gamma_{1}$ and $\gamma_{2}$ have increased
dispersion. In the homogeneous cases the elements of $\gamma$ are
independent and identically distributed, which leads to correlations
with identical marginal distributions. The three bivariate distributions
are also identical because the pair of correlations always have one
index in common. We amplified the variance of $\gamma_{1}=G_{12}$
and $\gamma_{3}=G_{23}$ in the heterogeneous case. From the contour
plots it is evident that the increased variance of the two elements
of $\gamma$ primarily increases the variance of the corresponding
correlations $C_{12}$ and $C_{23}$, whereas the effect on $C_{13}$
is modest.

\subsection{Additional Dependence Reduction}

The Jacobian is not perfectly diagonal and this partly explains the
dependence between the correlations, which can be seen in the contour
plots in the upper panels of Figure \ref{fig:heter_gamma}. We can
account for the structure in $J_{0}$ to reduce the dependence between
individual random correlations. From the Taylor expansion, $\varrho(\gamma)\approx\varrho(\gamma_{0})+J_{0}\cdot(\gamma-\gamma_{0}),$
it follows that $\mathrm{var}(\varrho(\gamma))\approx J_{0}\mathrm{var}(\gamma)J_{0}{}^{\prime}$.
Therefore, if we set $\mathrm{var}(\gamma)=\Sigma=J_{0}^{-1}\Lambda_{\omega}J_{0}^{-1}$,
then $\mathrm{var}(\varrho(\gamma))\approx\Lambda_{\omega}$. This
first-order approximation is reliable when $\Sigma=\mathrm{var}(\gamma)$
is small, whereas the nonlinearities in $C(\gamma)$ becomes important
if $\Sigma$ is large. 

The results based on $\gamma\sim N_{3}(\gamma_{0},J_{0}^{-1}\Lambda_{\omega}J_{0}^{-1})$,
with $\gamma_{0}=(\tfrac{1}{4},\tfrac{1}{4},\tfrac{1}{4})^{\prime}$
and $\Lambda_{\omega}=\text{diag}(\tfrac{\omega}{100},\tfrac{1}{100},\tfrac{\omega}{100})$
are presented in the lower panels of Figure \ref{fig:heter_gamma}.
The Jacobian and its inverse are (for this $\gamma_{0}$) given by,

\[
J_{0}=\left(\begin{array}{ccc}
0.920 & {\color{teal}0.102} & {\color{teal}0.102}\\
{\color{teal}0.102} & 0.920 & {\color{teal}0.102}\\
{\color{teal}0.102} & {\color{teal}0.102} & 0.920
\end{array}\right),\qquad\text{and}\quad J_{0}^{-1}=\left(\begin{array}{ccc}
1.111 & {\color{purple}-0.111} & {\color{purple}-0.111}\\
{\color{purple}-0.111} & 1.111 & {\color{purple}-0.111}\\
{\color{purple}-0.111} & {\color{purple}-0.111} & 1.111
\end{array}\right),
\]
respectively. The solid blue contour lines correspond to the homogeneous
case ($\omega=1$) and the red dashed contour lines correspond to
the heterogeneous case ($\omega=10$), as in the upper panels. It
is not possible to eliminate the dependence between the random correlations
entirely. However, the simple Jacobian-based adjustment does reduce
the linear dependence, which can be seen by comparing the contour
lines in the lower panels with those in the upper panels.

\subsection{A Bound for Smallest Eigenvalue of $C(\gamma)$}

The new method also makes it simple to bound the smallest eigenvalue
of the random correlation matrix, which avoids ill-conditioned matrices.
This can be done by bounding the range for the elements of $\gamma$. 
\begin{thm}
\label{thm:Bound}Let $\gamma_{\max}=\max_{k}|\gamma_{k}|$ be the
largest element of $\gamma$ in absolute value. Then,
\[
e^{-K\gamma_{\max}}\leq\lambda_{\min}\leq e^{-\gamma_{\max}},
\]
for some $K<\infty$. 
\end{thm}
The first inequality in Theorem \ref{thm:Bound} shows that the smallest
eigenvalue of $C(\gamma)$ is bounded away from zero by placing a
bound on $\max_{k}|\gamma_{k}|$, and we conjecture that $K=n$. Interestingly,
we note that $\exp(\text{\textminus}n\gamma_{\max})\simeq\tfrac{ne^{-n\gamma_{\max}}}{n-1+e^{-n\gamma_{\max}}}$
for large values of $\gamma_{\max}$, where the latter is the smallest
eigenvalue of an equicorrelation matrix with a common negative correlation.\footnote{This is the case where the common off-diagonal elements of $\log C$
equals $-\gamma_{\max}$, and we note that $-\gamma_{\max}=z(r)\rightarrow-\infty$
as $r\rightarrow-\tfrac{1}{n-1}$, see (\ref{eq:z(r)EquiC}).}

\begin{figure}[H]
\begin{centering}
\includegraphics[width=1\textwidth]{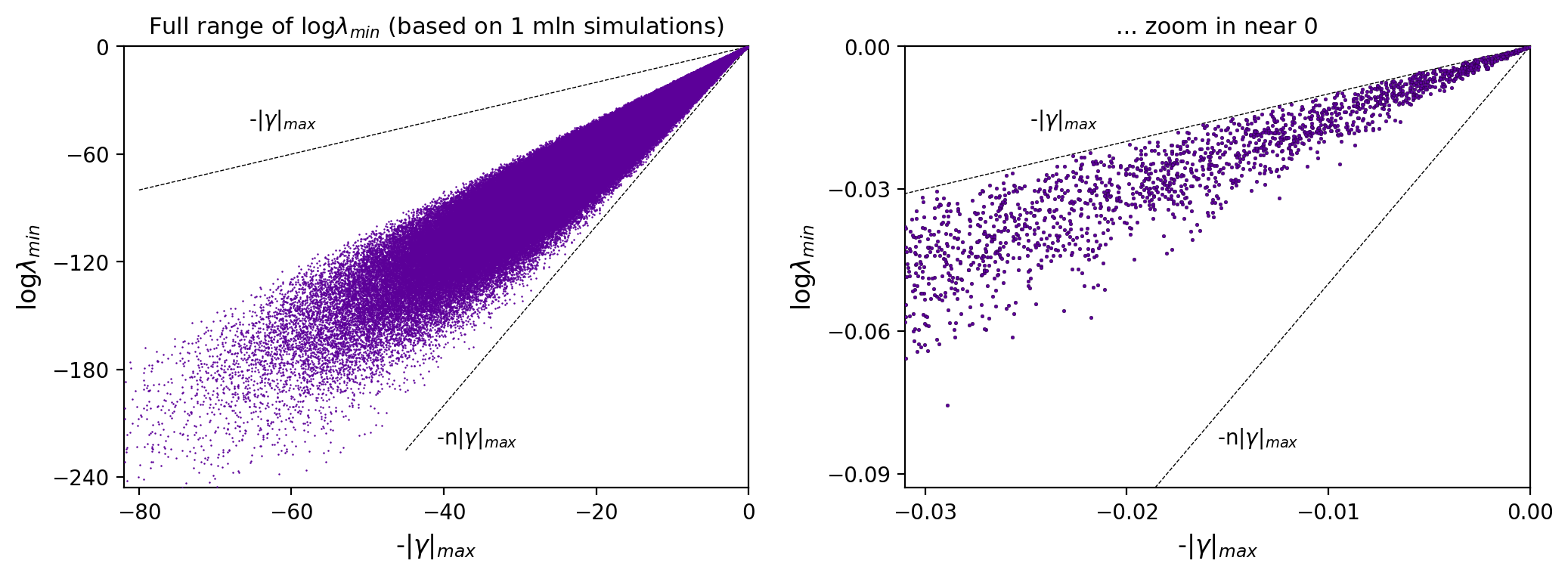}
\par\end{centering}
\caption{Scatter plots of $\log\lambda_{\min}$ against $-\gamma_{\max}$ for
one million random $5\times5$ correlation matrices.\label{fig:MinLambdaBounds}}
\end{figure}

In Figure \ref{fig:MinLambdaBounds} we have plotted $\log\lambda_{\min}$
against $-\gamma_{\max}$ for one million random correlation matrices
with dimension $n=5$ along with the conjectured upper and lower bound
for $\log\lambda_{\min}$. The lower bound appears to be binding for
very large values of $\gamma_{\max}$, whereas the upper bound only
becomes binding for $\gamma_{\max}\simeq0$. The latter corresponds
to the case where $C\simeq I$.

\subsection{\label{subsec:Resembling-the-Distribution}Resembling the Distribution
of Empirical Correlation Matrices}

The method can be used to approximate the distribution of empirical
correlation matrices. Let $\hat{C}$ be an empirical correlation matrix
computed from $T$ observations and consider $\hat{\gamma}=g(\hat{C})$.
Under suitable regularity conditions, \citet{ArchakovHansen:Correlation}
showed that $\sqrt{T}(\hat{\gamma}-\gamma)\overset{d}{\rightarrow}N(0,V_{\gamma})$
and derived an expression for $V_{\gamma}$. This asymptotic approximation
works well in finite samples and the off-diagonal elements of $V_{\gamma}$
tend to be close to zero, especially for high-dimensional correlation
matrices, see \citet{ArchakovHansen:GFTsimulations}. This suggests
that the new method can be used to resemble the distributions of empirical
correlation matrices by drawing $\gamma$ from a suitable Gaussian
distribution.

\section{Random Correlation Matrices with Special Structures}

\subsection{Non-Negative and Positive Random Correlation Matrices}

In this section, we show that non-negative correlations are guaranteed
if all elements of $\gamma$ are non-negative, and strictly positive
correlation are guaranteed if the elements of $\gamma$ are strictly
positive. The latter would, by the Perron-Frobenius theorem, ensure
that the eigenvector associated with the largest eigenvalue of $C$
had strictly positive elements. 

We borrow some terminology from the Markov chain literature for the
purpose of generating non-negative and positive correlation matrices.
\begin{defn}
An $n\times n$ matrix, $A$, is \emph{reducible} if there exists
a permutation matrix, $P$, such that $B=PAP^{\prime}=\left[\begin{array}{cc}
B_{[1,1]} & B_{[1,2]}\\
B_{[2,1]} & B_{[2,2]}
\end{array}\right]$ has $B_{[2,1]}=0_{n_{2}\times n_{1}}$, where $n_{1},n_{2}\geq1$
and $n_{1}+n_{2}=n$, otherwise $A$ is said to be \emph{irreducible}. 
\end{defn}
Because a correlation matrix is symmetric, it follows that $C=\mathrm{corr}(X)$
is reducible if and only if the variables can be reordered, $\tilde{X}=PX$,
such that $\tilde{C}=\mathrm{corr}(\tilde{X})$ is a block diagonal
matrix, i.e. 
\[
\tilde{C}=PCP^{\prime}=\left[\begin{array}{cc}
\tilde{C}_{[1,1]} & 0\\
0 & \tilde{C}_{[2,2]}
\end{array}\right],
\]
in which case we observe that
\[
\tilde{G}=\log\tilde{C}=\left[\begin{array}{cc}
\log\tilde{C}_{[1,1]} & 0\\
0 & \log\tilde{C}_{[2,2]}
\end{array}\right],
\]
has the same block diagonal structure. This shows that $C$ is reducible
if and only if $G$ is reducible. 
\begin{thm}
\label{thm:PositiveC}If $\gamma_{k}\geq0$ for all $k=1,\ldots,d$,
then all elements of $C(\gamma)$ are non-negative. Moreover, if $G(\gamma)=\log C(\gamma)$
is irreducible then all elements of $C(\gamma)$ are strictly positive.
\end{thm}
An implication of Theorem \ref{thm:PositiveC} is that $\gamma_{k}>0$
for all $k$ will translate to a $C(\gamma)$ with strictly positive
elements, because $G(\gamma)$ is irreducible in this case. 

It is worth mentioning that $\tilde{\gamma}\geq\gamma\geq0$ $\centernot\implies$
$C(\tilde{\gamma})\geq C(\gamma)$, as illustrated with the following
counterexample:
\[
C(\left[\begin{array}{c}
0.60\\
1.50\\
0.05
\end{array}\right])=\left[\begin{array}{ccc}
1 & 0.507 & 0.897\\
0.507 & 1 & 0.325\\
0.897 & 0.325 & 1
\end{array}\right],\qquad C(\left[\begin{array}{c}
0.59\\
0.50\\
0.04
\end{array}\right])=\left[\begin{array}{ccc}
1 & 0.528 & 0.460\\
0.528 & 1 & 0.166\\
0.460 & 0.166 & 1
\end{array}\right].
\]

\subsection{Equicorrelation Matrices\label{subsec:Equicorrelation-Matrices}}

An equicorrelation matrix, $C$, is a correlation matrix where all
the correlations are identical. The corresponding $\gamma=g(C)$ is
a vector whose elements have the same value. Let $r$ denoted the
common correlation coefficient in $C$ and let $z$ be the corresponding
common element of $\gamma$, then the relationship between the two
is given by,
\begin{equation}
z(r)=\frac{1}{n}\log\left(1+n\tfrac{r}{1-r}\right),\label{eq:z(r)EquiC}
\end{equation}
and the inverse transformation is $r(z)=\frac{1-e^{-nz}}{1+(n-1)e^{-nz}}$,
see e.g. \citet{ArchakovHansen:Correlation}. An equicorrelation matrix
has two eigenvalues, $1+r(n-1)$ and $1-r$, where the latter has
multiplicity $n-1$, see \citet{OlkinPratt:1958}. Thus, the $n\times n$
equicorrelation matrix is positive definite if and only if $r\in(-\tfrac{1}{n-1},1)$. 

The following theorem establishes a relationship between the Beta
distribution for $r$ and a generalized logistic distribution for
$z$.
\begin{thm}
\label{thm:Logistic2Beta}Let $\gamma=(z,\ldots,z)^{\prime}\in\mathbb{R}^{d}$
with $d=n(n-1)/2$. Then $C(\gamma)$ is an equicorrelation matrix,
where the common correlation coefficient, $r$,  is confined to the
interval $(-\tfrac{1}{n-1},1)$ for all $z\in\mathbb{R}$. Moreover,
if $z$ has density,
\begin{equation}
f_{z}(z)=\frac{1}{B(\alpha,\beta)}\frac{e^{-\beta\tfrac{z-\mu}{s}}}{s\left(1+e^{-\tfrac{z-\mu}{s}}\right)^{\alpha+\beta}},\qquad z\in\mathbb{R},\label{eq:LogisticType4ToBeta}
\end{equation}
where $\mu=\tfrac{\log(n-1)}{n}$ and \textup{$s=\tfrac{1}{n}$},
then $r$ is Beta distributed, $B(\alpha,\beta)$, on the interval
$(-\tfrac{1}{n-1},1)$.
\end{thm}
The density (\ref{eq:LogisticType4ToBeta}) was introduced in \citet{Prentice1975}
and is known as the Generalized Logistic Distribution of Type IV.
This distribution is also referred to as the Exponential Generalized
Beta distribution of the second type, see e.g. \citet{CaivanoHarvey2014}.

If we set $\alpha=\beta=1$, it follows immediately that $r$ is uniformly
distributed on $(-\tfrac{1}{n-1},1)$.
\begin{cor}
\label{Corr:Logistic2Uniform}Let $\gamma=(z,\ldots,z)^{\prime}\in\mathbb{R}^{d}$
with $d=n(n-1)/2$, and suppose that $z$ is logistically distributed,
\begin{equation}
f_{z}(z)=\frac{e^{-\tfrac{z-\mu}{s}}}{s\left(1+e^{-\tfrac{z-\mu}{s}}\right)^{2}},\qquad z\in\mathbb{R},\label{eq:LogisticToUniform}
\end{equation}
where $\mu=\tfrac{\log(n-1)}{n}$ and \textup{$s=\tfrac{1}{n}$},
then $r$ is uniformly distributed on the interval $(-\tfrac{1}{n-1},1)$.
\end{cor}
In the special case where $n=2$, we have $\mu=0$ and $s=1/2$ and
the logistic distribution in (\ref{eq:LogisticToUniform}) is also
known as a Fisher $z$-distribution with $(d_{1},d_{2})=(2,2)$ degrees
of freedom.\footnote{Moreover, in this case where $Z\sim\mathrm{logistic}(0,\tfrac{1}{2})$
we also have that $\exp(2Z)\sim F(2,2)$, (the $F$-distribution with
degrees of freedom $d_{1}=d_{2}=2$).} 

Theorem \ref{thm:Logistic2Beta} provides valuable insight about the
dispersion of the elements of $\gamma$ as the dimension of the correlation
matrix, $n$, increases. The variance for the density in (\ref{eq:LogisticToUniform})
is $\mathrm{var}(z)=\frac{\pi^{2}}{3}n^{-2}$. This suggests that
a scaling factor of $1/n$ should be used on the elements of $\gamma$
to preserve similar dispersion for the correlation coefficients in
$C(\gamma)$ as $n$ increases.

\subsection{Block Correlation Matrices}

If $C$ has a block structure then $\log C$ and $C^{-1}$ has the
same block structure, see \citet{ArchakovHansen:CanonicalBlockMatrix}.
This can be used to generate random correlation matrices with block
structures, as well as random precision matrices, $C^{-1}$, with
block structures, while positive definiteness is guaranteed. A correlation
matrix has a block structure if 
\[
C=\left[\begin{array}{cccc}
C_{[1,1]} & C_{[1,2]} & \cdots & C_{[1,K]}\\
C_{[2,1]} & C_{[2,2]}\\
\vdots &  & \ddots\\
C_{[K,1]} &  &  & C_{[K,K]}
\end{array}\right]\in\mathbb{R}^{n\times n},
\]
where the diagonal blocks, $C_{[i,i]}\in\mathbb{R}^{n_{i}\times n_{i}}$,
where $i=1,...,K$, have ones along the diagonal and $\rho_{i,i}\in(-1,1)$
in all off-diagonal elements (i.e., equicorrelation structure), and
the off-diagonal blocks, $C_{[i,j]}\in\mathbb{R}^{n_{i}\times n_{j}}$,
where $i,j=1,\ldots,K$ and $i\neq j$, $n_{1}+\cdots+n_{K}=n$ have
all elements equal to $\rho_{i,j}\in(-1,1)$. Symmetry is guaranteed
with $\rho_{i,j}=\rho_{j,i}$.The values $\rho_{i,j}$ must also be
such that $C$ is a positive definite matrix.

A useful property of this structure is that the matrix logarithm,
$G=\log C$, is also a block matrix with the same block structure
as $C$. Thus, 
\begin{equation}
G_{[k,k]}=\left[\begin{array}{cccc}
y_{k} & \gamma_{k,k} & \cdots & \gamma_{k,k}\\
\gamma_{k,k} & y_{k} & \ddots & \vdots\\
\vdots & \ddots & \ddots & \gamma_{k,k}\\
\gamma_{k,k} & \cdots & \gamma_{k,k} & y_{k}
\end{array}\right]\in\mathbb{R}^{n_{k}\times n_{k}}\qquad G_{[k,l]}=\left[\begin{array}{ccc}
\gamma_{k,l} & \cdots & \gamma_{k,l}\\
\vdots & \ddots & \vdots\\
\gamma_{k,l} & \cdots & \gamma_{k,l}
\end{array}\right]\in\mathbb{R}^{n_{k}\times n_{l}},\quad k\neq l,\label{eq:BlockG}
\end{equation}
with $\gamma_{k,l}\in\mathbb{R}$ and $\gamma_{k,l}=\gamma_{l,k}$
for $i,j=1,\ldots,K$. Matrix $C$ is uniquely determined from the
off-diagonal elements of $G$, and the inverse mapping can be obtained
with the algorithm in \citet{ArchakovHansen:Correlation}. The problem
is to determine a $n\times1$ diagonal vector for $G$ such that $\exp\{G\}$
is a correlation matrix. Generally, it requires the matrix exponential
to be evaluated for an $n\times n$ matrix (several times) and the
computational burden of this is of order $\mathcal{O}(n^{3}\log n)$.
For block matrices, the entries on the main diagonal are identical
within each diagonal block, so we have to determine only $K$ diagonal
elements, $y=(y_{1},...,y_{K})^{\prime}$, which greatly simplifies
the computational burden.

The matrix $G$ can be represented as $G=QDQ^{\prime}$, where $Q$
is an orthonormal matrix, $Q^{\prime}Q=I_{n}$, which does not depend
on the elements of $C$ (nor $G$). The corresponding closed-form
expression for $D$ is
\begin{equation}
D=\left[\begin{array}{cccc}
A+\mathrm{diag}(y) & 0 & \cdots & 0\\
0 & (y_{1}-\gamma_{1,1})I_{n_{1}-1} &  & \vdots\\
\vdots &  & \ddots & 0\\
0 & \cdots & 0 & (y_{K}-\gamma_{K,K})I_{n_{K}-1}
\end{array}\right],\label{eq:Dmatrix}
\end{equation}
where $A$ is a $K\times K$ matrix with elements,
\[
A_{k,l}=\begin{cases}
\gamma_{k,k}(n_{k}-1) & \text{for }k=l,\\
\gamma_{k,l}\sqrt{n_{k}n_{l}} & \text{for }k\neq l,
\end{cases}
\]
and $\mathrm{diag}(y)$ is the $K\times K$ diagonal matrix with the
elements of $y$ along the diagonal, see \citet{ArchakovHansen:CanonicalBlockMatrix}
for details. In this representation, all distinct off-diagonal entries
of $G$ appear in $K\times K$ upper left diagonal block of $D$.
This is convenient, as it can be shown that to restore the original
matrix $C$ from given values $\gamma_{k,l}$, we only need to find
a proper vector $y$ which determines the diagonal of this block as
well as the entire main diagonal of $D$. 
\begin{thm}
\label{thm:Block}Let $G$ be of the form (\ref{eq:BlockG}) for some
$n_{1},\ldots,n_{K}\in\mathbb{N}$. Given any constants, $\gamma_{k,l}\in\mathbb{R}$,
$1\leq k,l\leq K$, with $\gamma_{k,l}=\gamma_{l,k}$, there exist
unique constants, $y_{1}^{\ast},\ldots,y_{K}^{\ast}\leq0$, such that
$\exp G[y]$ is a block correlation matrix. The unique $y^{\ast}$
can be determined by iterating on,
\[
y_{k}^{(N+1)}=y_{k}^{(N)}+\log n_{k}-\log\left([\exp\{A+\mathrm{diag}(y^{(N)})\}]_{kk}+(n_{k}-1)\,e^{y_{k}^{(N)}-\gamma_{k,k}}\right),
\]
until convergence from an arbitrary starting value, $y^{(0)}\in\mathbb{R}^{K}$. 
\end{thm}
The computational burden of this algorithm is of order $\mathcal{O}(K^{3}\log K)$,
which is a substantial simplification relative to the generic algorithm
in \citet{ArchakovHansen:Correlation} whenever $K$ is smaller than
$n$.\footnote{For a $200\times200$ block correlation matrix with $K=10$ blocks,
the contraction is about 175 times faster than the generic algorithm,
which does not take advantage of the block structure, and reduce the
memory requirements by a factor of about 30.} Theorem \ref{thm:Block} shows that in order to generate a random
block correlation matrix, it suffices to generate the off-diagonal
entries of $G$, $\gamma_{k,l}\in\mathbb{R}$, and then recover the
unique vector $y^{\ast}\in\mathbb{R}^{K}$. The algorithm in Theorem
\ref{thm:Block} ensures that $\exp G[y^{*}]$ has ones along the
main diagonal and is a valid block-correlation matrix. Moreover, all
elements of $C=\exp G[y^{*}]$ are available in a closed-form as functions
of $\gamma_{k,l}$ and $y^{*}$. An evaluation of matrix exponential
for the $n\times n$ matrix $G[y^{*}]$ is not needed.

It is straight forward to generate random block correlation matrices
in the vicinity of a particular block correlation matrix using the
method described here, and it is obviously also possible to generate
random correlation matrices (without a block structure) in the vicinity
of a particular block correlation matrix using the standard algorithm
proposed in \citet{ArchakovHansen:Correlation}.

\subsubsection{Random correlation matrices of (very) large dimensions}

\begin{figure}[p]
\begin{centering}
\includegraphics[width=0.85\textwidth]{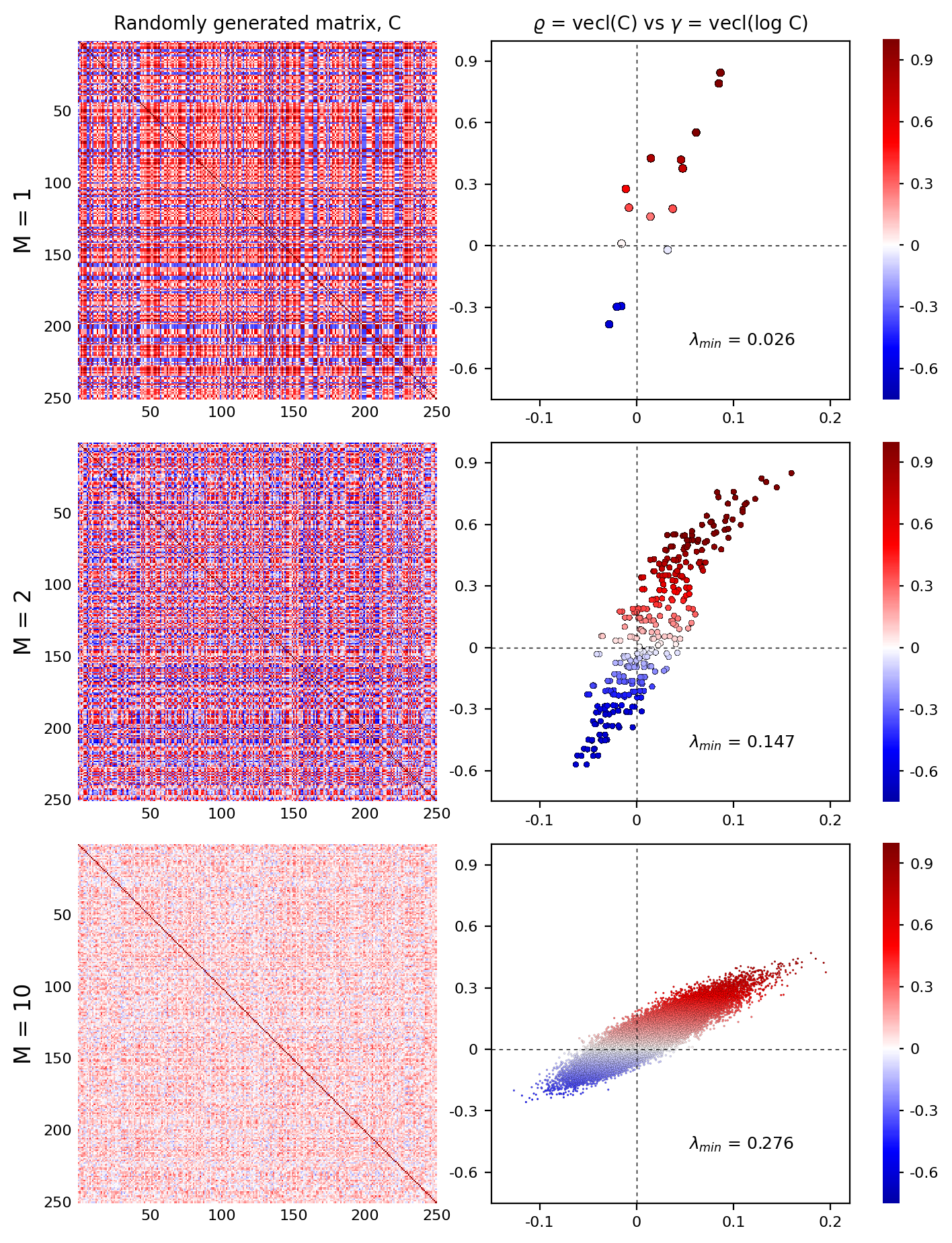}
\par\end{centering}
\caption{The left side presents random $250\times250$ correlation matrices,
$C$, constructed as the average of 1, 2 and 10 random block matrices,
whose rows and columns were subject to random permutations. The right
side presents scatter plots with the corresponding correlation elements
contained in the resulting matrices ($\varrho=\text{vecl}(C)$ are
drawn against $\gamma=\text{vecl}(\log C)$). \label{fig:A-random-250x250}}

\end{figure}

The canonical representation of block matrices can also be used to
efficiently generate high-dimensional correlation matrices by taking
convex combinations of permutated random block matrices, i.e., 
\[
C=\sum_{m=1}^{M}\omega_{m}P_{m}Q\exp\{D_{m}\}Q^{\prime}P_{m}^{\prime},\qquad\sum_{m}\omega_{m}=1,\quad\omega_{m}\geq0,
\]
where $D_{m}$, $m=1,\ldots,M$ are constructed from random $\gamma_{k,l}$,
$1\leq k\leq l\leq K$ with the block structure (\ref{eq:Dmatrix})
and $P_{m}$ are perturbation matrices.\footnote{In this case, the computational burden is of order $\mathcal{O}(M\times K^{3}\log K)$.}
Figure \ref{fig:A-random-250x250} presents random $250\times250$
correlation matrices, which are constructed from $M=1$ (upper plots),
$M=2$ (middle plots) and $M=10$ (bottom plots) random block correlation
matrices, each having $5\times5$ blocks (each block is of size $50\times50$),
such that each block correlation matrix has 15 distinct correlation
coefficients. Before averaging the matrices, the rows (and columns)
are shuffled with random perturbations. The resulting matrices are
guaranteed to be positive definite, and the corresponding smallest
eigenvalues are also reported in the Figure. As we can observe, the
generated random matrix fastly departs from the block structure as
$M$ increases, which is manifested by the diversity of the corresponding
correlation elements rising quickly with $M$.

\section{Existing Methods for Generating Random Correlation Matrices}

There is a large literature on generating random correlation matrices,
see \citet{MarsagliaOlkin:1984} and \citet{Pourahmadi2011} for references.
In this section, we discuss some existing methods for generating random
correlation matrices, and compare some of their features and properties
with those of the new method. 

\subsection{Naive Method}

A simple method to generate random correlation matrices is to simply
generate random correlation coefficients, $C_{i,j}\in[-1,1]$, $1\leq i<j\leq n$,
set $C_{ii}=1,$ for $i=1,\ldots n$, $C_{i,j}=C_{j,i}$ for $i<j$,
and then discard the invalid correlation matrices, which are characterized
by $\lambda_{\min}(C)<0$.

\begin{figure}[H]
\begin{centering}
\includegraphics[width=0.9\textwidth]{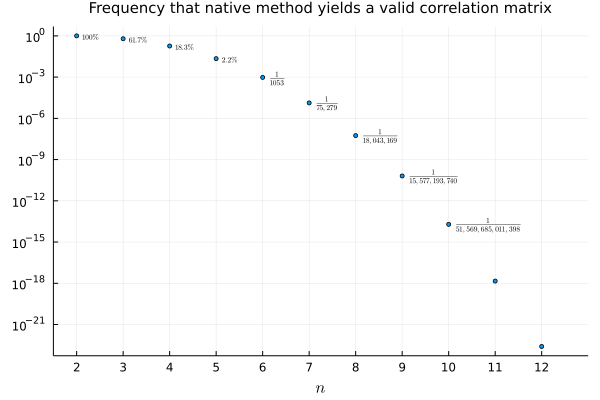}
\par\end{centering}
\caption{The probability that a symmetric matrix with independent and uniformly
distributed elements, $C_{ij}$, $1\protect\leq i<j\protect\leq n$
on $[-1,1]$, and $C_{ii}=1$, $i=1,\ldots,n$, is a valid correlation
matrix.\label{fig:NaiveMethod}}

\end{figure}

This approach yields a uniform distribution over the set of valid
correlation matrices when $C_{ij}$, $1\leq i\leq j\leq n$ are independent
and uniformly distributed on $[-1,1]$. Interestingly, the correlation
coefficients, $C_{i,j}$, in the retained correlation matrices are
beta distributed on $[-1,1]$, $B(\alpha,\alpha)$ with $\alpha=n/2$.
This can be inferred from results in \citet{Joe:2006}. This naive
method for generating random correlation matrices is very inefficient
and impractical except for very low dimensional matrices. With $n=6$
the percentage of matrices with negative eigenvalues is more than
99.9\%, and for $n=10$ it takes about 55 quadrillions random matrices
to get a single valid correlation matrix, see Figure \ref{fig:NaiveMethod}.
This approach clearly impractical except for small $n$. 

\subsection{Random Gram Methods}

A valid correlation matrix can be obtained from any $m\times n$ matrix,
$U=(u_{1},\ldots,u_{n})$, with normalized columns, $u_{j}^{\prime}u_{j}=1$,
for $j=1,\ldots,n$. It follows immediately that $C=U^{\prime}U$
is positive semidefinite with ones along the diagonal, and if $U$
has rank $n$, then $C=U^{\prime}U$ is a non-singular correlation
matrix. Several methods are based on this idea (typically with $m=n$),
where a random correlation matrix is obtained from random vectors,
$u_{1},\ldots,u_{n}$, on the unit sphere, $S_{m}=\{u\in\mathbb{R}^{m},u^{\prime}u=1\}$.
The random Gram method generates $n$ vectors on $S_{n}$ and the
Gram matrix $C=U'U$ is the resulting random correlation matrix. The
uniform distribution on $S_{n}$ was discussed in \citet{MarsagliaOlkin:1984},
see also \citet{Holmes:1991}, and it generates a $C$ where the marginal
distributions of the correlation coefficients are Beta distributed,
$B(\frac{1}{2},\frac{n-1}{2})$. The vectors, $u_{j},$ $j=1,\ldots,n$,
can be drawn from other distributions, such as those proposed by \citet{TuitmanVanduffelYao:2020},
which ensures that the average correlation coefficient is centered
about a particular value.

\subsection{Standard Angles Parameterization (SAP) Method}

A variant of the Random Gram method is the case where $U$ is a triangular
matrix. This choice was discussed in \citet{MarsagliaOlkin:1984}
and a particular triangular form was proposed by \citet{PinheiroBates:1996}.
Their choice for $U$ is defined by the angles, $\theta_{ij}\in[0,\pi)$,
for $1\leq i<j\leq n$, such that 

\[
U=\left[\begin{array}{cccccc}
1 & \cos\theta_{1,2} & \cos\theta_{1,3} & \cdots & \cos\theta_{1,n-1} & \cos\theta_{1,n}\\
0 & \sin\theta_{1,2} & \cos\theta_{2,3}\sin\theta_{1,3} & \cdots & \cos\theta_{2,n-1}\sin\theta_{1,n-1} & \cos\theta_{2,n}\sin\theta_{1,n}\\
0 & 0 & \Pi_{i=1}^{2}\sin\theta_{i,3} &  & \cos\theta_{3,n-1}\Pi_{i=1}^{2}\sin\theta_{i,n-1} & \cos\theta_{3,n}\Pi_{i=1}^{2}\sin\theta_{i,n}\\
\vdots & \vdots & \ddots & \ddots &  & \vdots\\
0 & 0 & 0 &  & \Pi_{i=1}^{n-2}\sin\theta_{i,n-1} & \cos\theta_{n-1,n}\Pi_{i=1}^{n-2}\sin\theta_{i,n}\\
0 & 0 & 0 & \cdots & 0 & \Pi_{i=1}^{n-1}\sin\theta_{i,n}
\end{array}\right]
\]
is an upper triangular matrix. This requires $d=n(n-1)/2$ angles,
$\theta_{ij}$, and it follows that any distribution on $[0,\pi)^{d}$
will correspond to some distribution over the space of correlation
matrices. If the angles are independent and uniformly distributed
on $[0,\pi)$, then $C$ has coefficients with very heterogeneous
marginal distributions. If one instead specifies $\theta_{ij}$ to
have the density
\[
f_{j}(x;\alpha)=\frac{\sin^{2\alpha-j}(x)}{B(\alpha-\frac{j-1}{2},\frac{1}{2})},\qquad j=1,\dots,n-1,
\]
for some $\alpha\geq n/2$, then marginal distributions of the correlation
coefficients are identical and Beta distributed, $\mathrm{Beta}(\alpha,\alpha)$
on the interval $[-1,1]$, see \citet{PourahmadiWang:2015}. This
is known as the \emph{Standard Angles Parameterization} (SAP) method.

\subsection{Eigendecomposition Method}

One of the first ways to generate random correlation matrices, see
\citet{Chalmers:1975} and \citet{BendelMickey:1978}, was based on
the eigendecomposition of the correlation matrix, $C=Q\Lambda Q^{\prime}$
where $Q^{\prime}Q=I$ and $\Lambda=\mathrm{diag}(\lambda_{1},\ldots,\lambda_{n})$. 

The premise of this method is a distribution of eigenvalues on the
$n$-simplex: $\{(\lambda_{1},\ldots,\lambda_{n}):\sum_{j}\lambda_{j}=n,\text{ }\lambda_{j}\geq0\}$.
Given a set of random eigenvalues, the method proceeds to determine
a set of eigenvectors (the columns of $Q$), such that $Q\Lambda Q^{\prime}$
is a valid correlation matrix. The latter is not a trivial step, because
the set of $Q$ matrices that produce a valid correlation matrix for
a given set of eigenvalues has measure zero in the set of all orthonormal
matrices. For the pair $(\Lambda,Q)$ to generate a valid correlation
matrix, the following conditions must be satisfied. 
\begin{enumerate}
\item The diagonal matrix, $\Lambda$, must satisfy $\lambda_{j}\geq0$,
$j=1,\ldots,n$, and $\sum_{j=1}^{n}\lambda_{j}=n$.
\item The matrix $Q=(q_{1},\ldots,q_{n})$ must be orthonormal, $q_{j}^{\prime}q_{j}=1$
and $q_{i}^{\prime}q_{j}=0$ for all $i\neq j=1,\ldots,n$. 
\item Combined they must satisfy $\mathrm{diag}(Q\Lambda Q^{\prime})=(1,\ldots,1)^{\prime}$
. 
\end{enumerate}
The last condition is a cross restriction on $\Lambda$ and $Q$.
Among all $Q$-matrices that satisfy the second condition, the fraction
of matrices that also satisfy the third condition for a particular
$\Lambda$, is zero. A method for determining a valid $Q$-matrix
is therefore needed, and such algorithms are given in \citet{Chalmers:1975},
\citet{BendelMickey:1978}, \citet{MarsagliaOlkin:1984}, and \citet{DaviesHigham:2000}.\footnote{\citet{Holmes:1991} provides a comprehensive study of the statistical
properties of spectral functions of correlation matrices generated
by Bendel and Mickey's algorithm. For financial applications, \citet{HuttnerMai:2019}
adapt the Bendel-Mickey Algorithm to generate correlation matrices
with a Perron-Frobenius property.} These methods begin with an initial (random) orthonormal matrix,
$Q_{0}$, that is subjected to successive transformations until a
valid $Q$-matrix is determined. The method by \citet{DaviesHigham:2000}
is implemented in the MATLAB function gallery('\texttt{randcorr}'). 

\subsection{Partial Correlations (PAC) Method}

The partial correlation (PAC) method by \citet{Joe:2006} uses random
partial correlations to generate random correlation matrices. Specifically
the $n(n-1)/2$ partial correlations given by 
\[
\varrho_{ij}=\frac{C_{ij}-d_{ij}^{(i,j)}}{\sqrt{(1-d_{ii}^{(i,j)})(1-d_{jj}^{(i,j)})}},\qquad\text{for}\quad1\leq i<j\leq n,
\]
where $d_{ij}^{(i,j)}=C_{i,I_{ij}}[C_{I_{ij},I_{ij}}]^{-1}C_{I_{ij},j}$,
and $C_{I_{ij},I_{ij}}=[C_{l,m}]_{i<l,m<j}$, $C_{i,I_{ij}}=[C_{i,m}]_{i<m<j}$,
and $C_{I_{ij},j}=C_{j,I_{ij}}^{\prime}$, are sub-matrices of $C$.
When $j=i+1$ the partial correlation is simply the correlation, $\varrho_{i,i+1}=C_{i,i+1}$;
otherwise, $\varrho_{ij}$ is the partial correlation between the
$i$-th and $j$-th variables, conditional on all variables indexed
between $i$ and $j$. Clearly any correlation matrix, $C$, will
map to $\{\varrho_{i,j}\}_{1\leq i<j\leq n}$ and any set of these
partial correlation in $(-1,1)$ will translate to a valid correlation
matrix. This is similar to the result for stationary time series derived
in \citet{Barndorff-NielsenSchou1973}. \citet{LewandowskiKurowickaJoe:2009}
builds on \citet{Joe:2006} to propose computationally fast ways to
generate high-dimensional random correlation matrices.

Interestingly, the determinant of $C$ is given by $\det C=\prod_{1\leq i<j\leq n}(1-\varrho_{ij}^{2})$,
see \citet[theorem 1]{Joe:2006}.\footnote{We have here simplified the expression \citet[theorem 1]{Joe:2006},
which involved three products over three indices.} The PAC method draws from a distribution on $(-1,1)^{d}$, with $d=n(n-1)/2$,
and reconstructs the correlations from the partial correlations. 

When the partial correlations, $\{\varrho_{i,j}\}_{1\leq i<j\leq n}$,
are drawn independently and from the Beta distribution, $\mathrm{Beta}(\alpha_{ij},\alpha_{ij})$
on $(-1,1)$, with $\alpha_{ij}=\alpha+(1-j+i)/2$, then the correlation
coefficients are identically distributed with $C_{ij}\sim\mathrm{Beta}(\alpha,\alpha)$,
where $\alpha>(n-2)/2$, see \citet{Joe:2006}. Moreover, the joint
density of all correlations becomes proportional to the determinant
of the correlation matrix to the power $\alpha-n/2$.\footnote{The notation in \citet{Joe:2006} is $\alpha_{ij}=a+(n-1-j+i)/2$
and $\alpha=a+(n-2)/2$, which we have modified to make the resulting
distribution directly comparable to the SAP method.}%
{} It follows that by setting $\alpha=n/2$, this method will generate
the same distribution as the naive method. 
\begin{figure}[H]
\begin{centering}
\subfloat[Random Gram]{\begin{centering}
\includegraphics[width=0.25\textwidth]{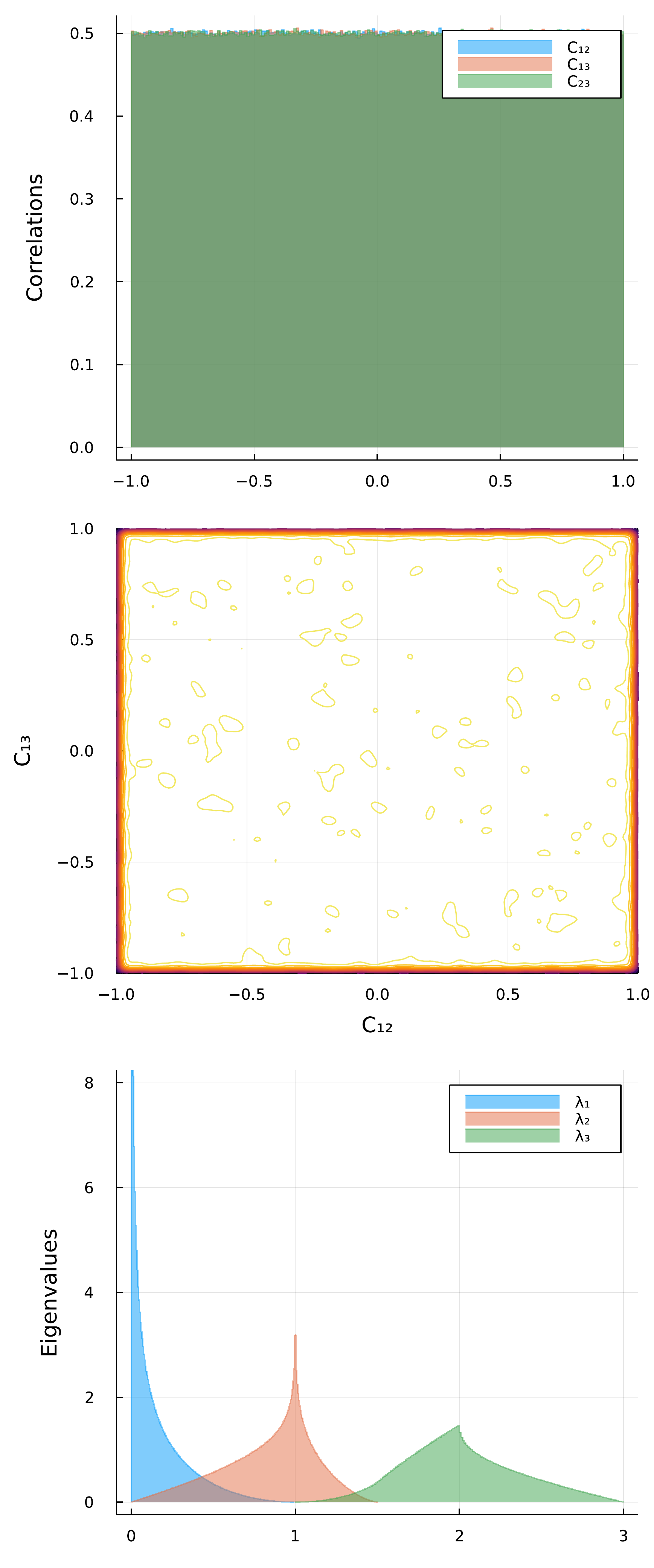}
\par\end{centering}
}\subfloat[PAC/SAP ($\alpha=1$)]{\centering{}\includegraphics[width=0.25\textwidth]{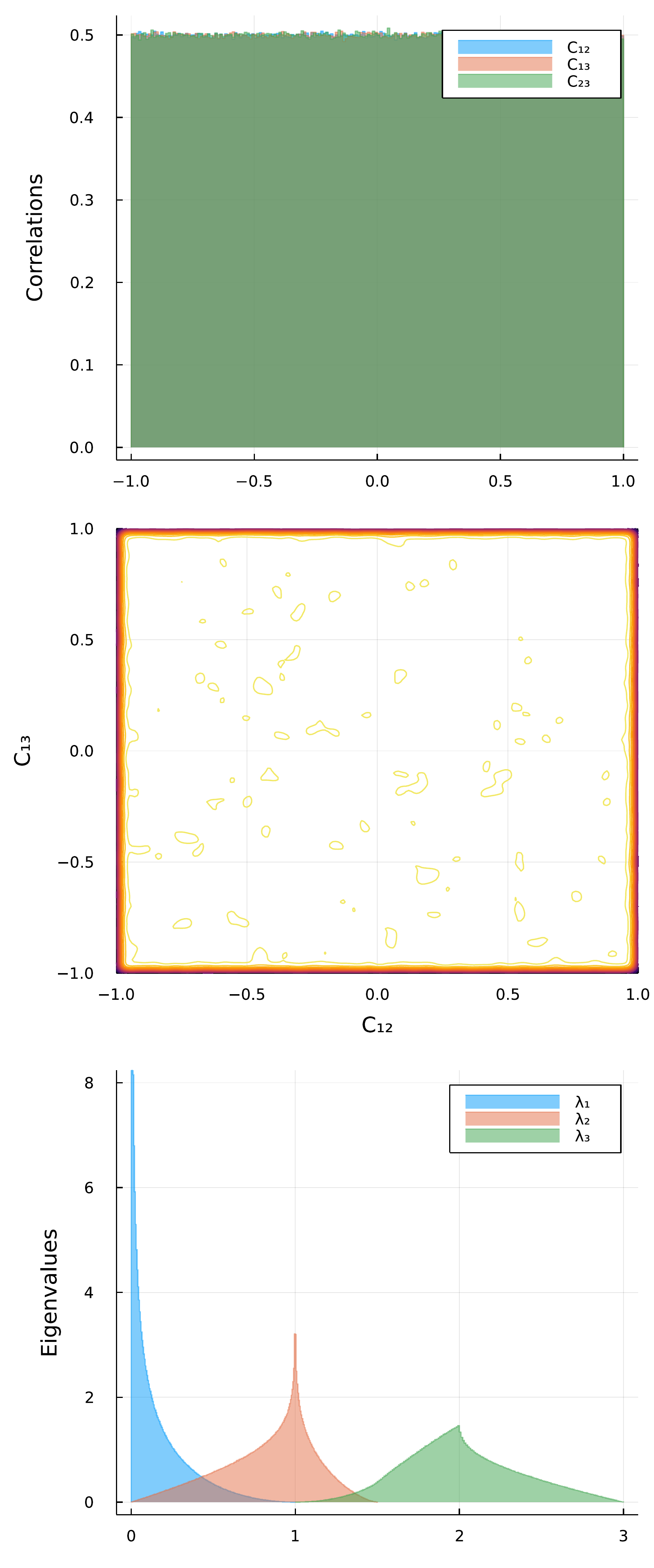}}\subfloat[PAC/SAP ($\alpha=10$)]{\centering{}\includegraphics[width=0.25\textwidth]{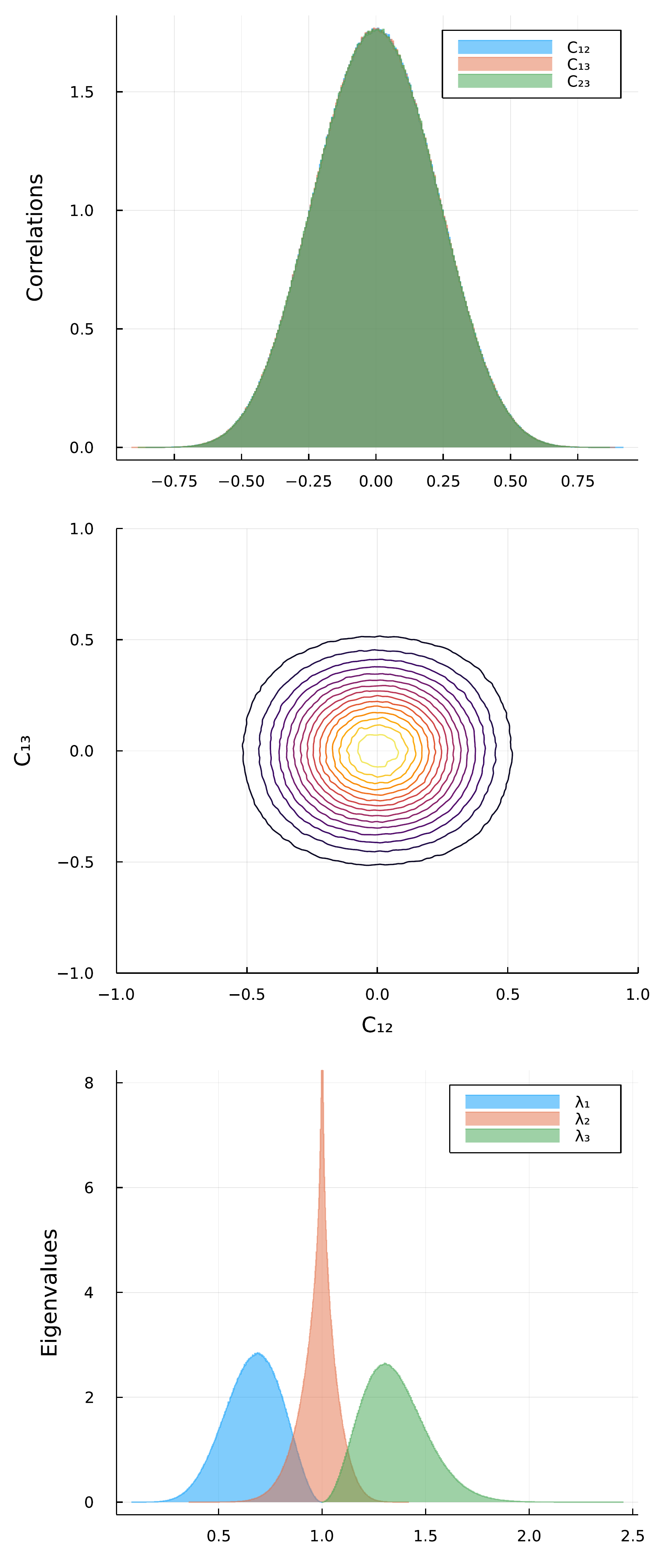}}\subfloat[Eigendecomposition]{\centering{}\includegraphics[width=0.25\textwidth]{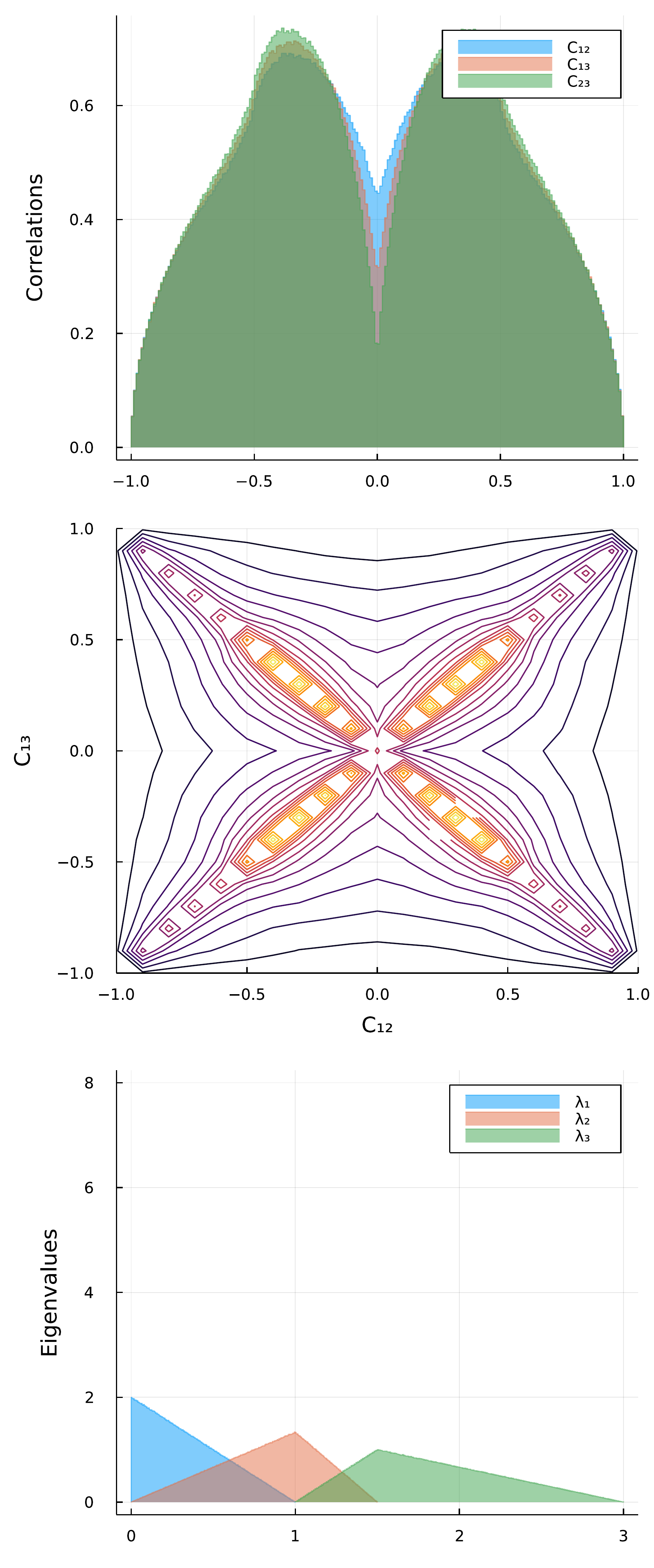}}
\par\end{centering}
\caption{{\small{}Properties of four existing methods for generating random
correlation matrices for the case $n=3$. (a) Random Gram with uniform
distribution on the sphere; (b,c) SAP and PAC methods with $\alpha=1$;
(d) Bendel and Mickey's method with eigenvalues $\lambda_{i}=y_{i}/(y_{1}+y_{2}+y_{3})$,
$i=1,\ldots,3$, with $y_{i}$ iid and distributed as Exp(1). Marginal
distributions are show for $C_{12}$, $C_{13}$, and $C_{23}$ (top)
and $\lambda_{1}$, $\lambda_{2}$, and $\lambda_{3}$(bottom), and
contour plots for bivariate density of $(C_{12},C_{13})$ is shown
in the middle panels.}\label{fig:dim3_density}}
\end{figure}

\begin{figure}[H]
\begin{centering}
\includegraphics[width=1\textwidth]{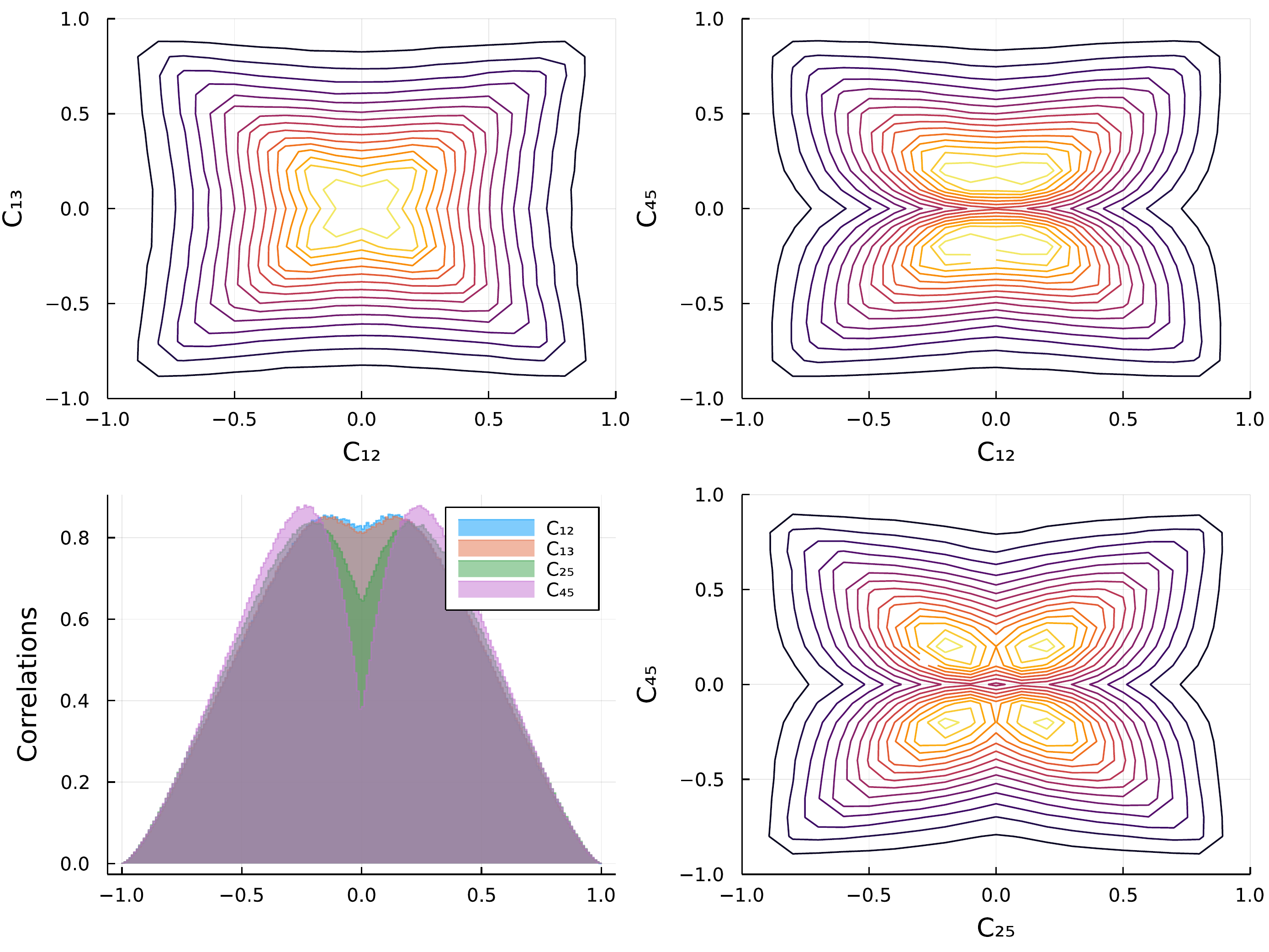}
\par\end{centering}
\caption{{\small{}Properties of eigendecomposition-based method for generating
random correlation matrices when $n=5$. Marginal distributions for
four correlations are shown in the lower-left panel and the three
other panels display contour plots for three bivariate distributions.}\label{fig:EigenBasedMethod}}
\end{figure}

The properties of some random correlation matrices, $n=3$, are shown
in Figure \ref{fig:dim3_density}. Panel (a) is the Random Gram method
where $u_{j}$, $j=1,\ldots,3$ are independent and uniformly distributed
on the sphere, $S_{3}$. This choice yields uniformly distributed
correlation coefficients when $n=3$. Panels (b) and (c) are the distributions
that the SAP and PAC methods produce with $\alpha=1$ and $\alpha=10$,
respectively. Panel (d) is the eigendecomposition-based method and
it produces rather bizarre marginal and joint distributions for the
correlations. This suggests that the algorithm used to determine a
valid orthonormal matrix, $Q$, results in some unexpected patterns
in the distribution for $C$. The marginal distributions are heterogeneous
and there are odd dependencies between correlation coefficients. We
have investigated this aspect of the eigendecomposition-based method
for $n=5$ in Figure \ref{fig:EigenBasedMethod}. It also shows very
heterogeneous and bimodal marginal distributions and rather bizarre
and heterogeneous contour plots for pairs of correlations, including
multimodal joint distributions.

\subsection{Random Correlations from Matrix Distributions}

Another popular approach for generating random covariance and correlation
matrices is based on the Wishart distribution and, more generally,
the matrix Gamma distribution. This method, which we will refer to
as the Wishart method, is frequently used in a Bayesian context. The
Wishart distribution is defined over symmetric positive semi-definite
matrices and arises as the distribution of a scaled sample covariance
matrix obtained from a sample of Normal random vectors. For instance,
if $X_{t}\sim\mathrm{iid}N_{n}(0,\Sigma)$, $t=1,...,T$, then $S=\sum_{t=1}^{T}X_{t}X_{t}^{\prime}$
is Wishart distributed with parameters $\Sigma$ and $T$, written
$S\sim\mathcal{W}_{n}(\Sigma,T)$, where $T$ is the degrees of freedom
parameter. We have that $\frac{1}{T}S$ is a sample covariance matrix,
and the corresponding sample correlation matrix is $\hat{C}=D_{S}^{-1}SD_{S}^{-1}$,
where $D_{S}=\text{diag}\bigl(S_{11}^{\frac{1}{2}},...,S_{nn}^{\frac{1}{2}}\bigl)$.
Thus, generating a random correlation matrix from the Wishart distribution,
$\mathcal{W}_{n}(\Sigma,T)$, is equivalent to computing a sample
correlation matrix, $\hat{C}$, from a random sample, $X_{t}\sim\mathrm{iid}N_{n}(0,\Sigma)$,
$t=1,...,T$. For $\hat{C}$ to be non-singular, the sample size,
$T$, must be at least as large as the matrix dimension, $n$. 

Generating a random correlation matrix in a vicinity of a target correlation
matrix, $C$, is possible with the Wishart method. This can be done
using $S=\sum_{t=1}^{T}X_{t}X_{t}^{\prime}$ where $X_{t}\sim\mathrm{iid}N_{n}(0,C)$,
$t=1,...,T$. However, there are some drawbacks to the Wishart method.
First, the possible range of dispersions for the individual correlations
is severely limited by the constraint: $T\geq n$. We have $\sqrt{T}(\hat{C}_{ij}-C_{ij})\overset{d}{\rightarrow}N(0,(1-C_{ij}^{2})^{2})$,
as $T\rightarrow\infty$, such that for large $T$, the random correlation
$\hat{C}_{ij}$ will be approximately distribution as $N(C_{ij},\frac{(1-C_{ij}^{2})^{2}}{T})$,
which shows that $\mathrm{var}(\hat{C}_{ij})$ is (approximately)
bounded to be below $\frac{1}{n}(1-C_{ij}^{2})^{2}$. Another implication
is that it is not possible to control the relative dispersion of different
elements of $C$ with the Wishart method; their variance is given
from $C$ and $T$, and their relative variance is asymptotically
determined from $C$ alone. 

The new method for generating random correlation matrices makes it
possible emulate the Wishart method. This is achieved with a single
random vector, $\gamma$, drawn from the appropriate Gaussian distribution,
see Section \ref{subsec:Resembling-the-Distribution}. An advantage
of the new method is that it is not bounded by the limitations of
the Wishart method, and the new method makes it simple to control
the relative dispersion of elements in $C$, as discussed in Section
\ref{subsec:Heterogenous-Marginal-Distributi}.

\begin{figure}[h]
\begin{centering}
\includegraphics[width=1\textwidth]{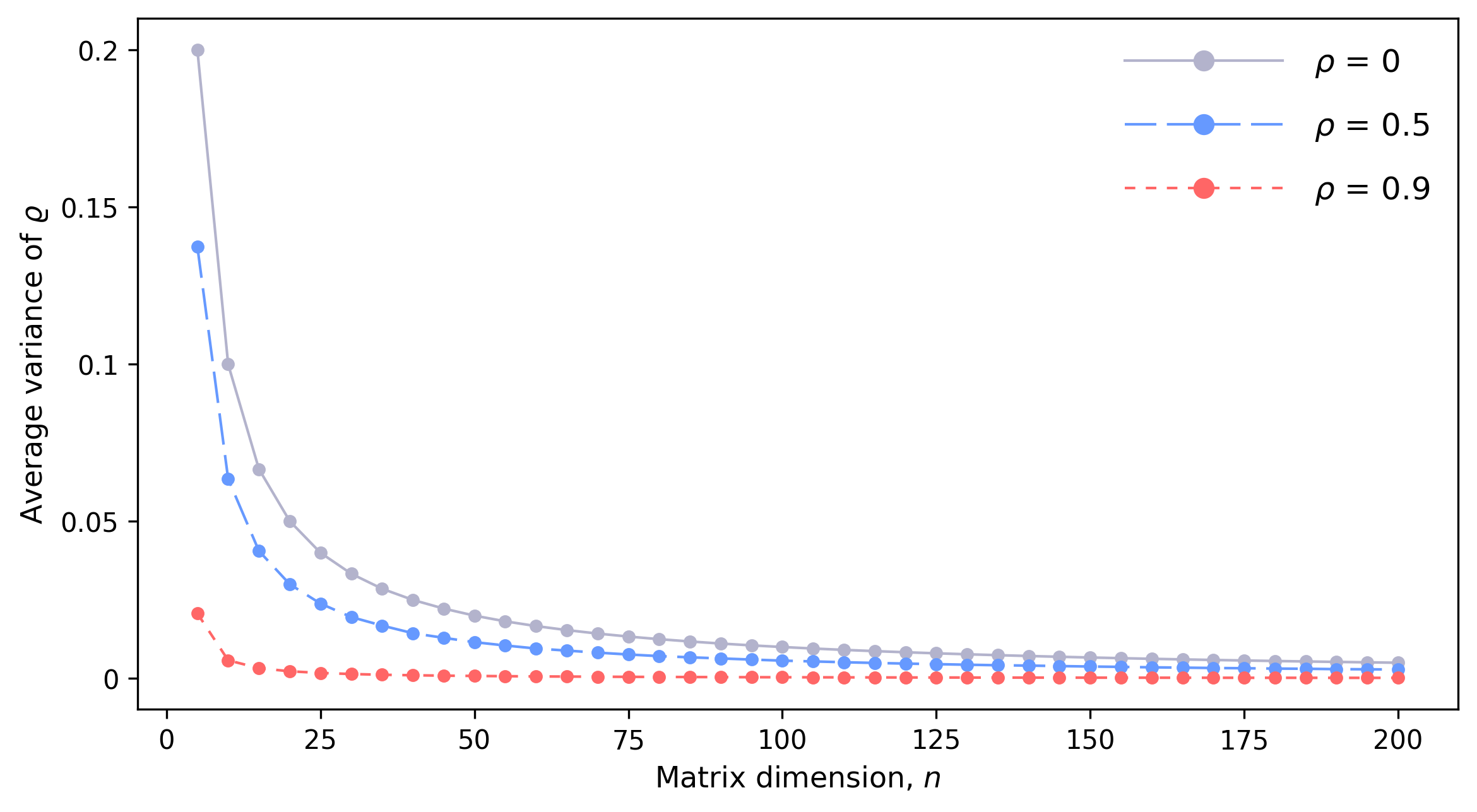}
\par\end{centering}
\caption{Average variances of elements in random correlation matrices generated
by the Wishart method, $\mathcal{W}_{n}(C,n)$, as a function of the
dimension $n$, where the target matrix, $C$, is an equicorrelation
matrix with $\rho=0.0$, $0.5$, and $0.9$. \label{fig:wishart_max_var}}
\end{figure}

The limitations of the Wishart method is illustrated in Figure \ref{fig:wishart_max_var}.
For a range of matrix dimensions, $n$, we generate random Wishart
correlation matrices with $T=n$, which corresponds to the largest
possible dispersion of the random correlations. For the target correlation
matrix we use equicorrelation matrices with $\rho=0$, $\rho=0.5$,
and $\rho=0.9$. Figure \ref{fig:wishart_max_var} presents the variance
of random correlation coefficients using the Wishart method for the
three target matrices. The upper bound for the variance drops rapidly
as $n$ increases, especially for $\rho=0.9$. 

\section{Summary}

In this paper, we have introduced a new method for generating random
correlation matrices. The method is based on a one-to-one mapping
between the space of non-singular correlation matrices, $\mathcal{C}_{n\times n}$,
and the space of real vectors, $\mathbb{R}^{d}$, where $d=n(n-1)/2$.
Any distribution on $\mathbb{R}^{d}$ translates to a distribution
on $\mathcal{C}_{n\times n}$ (and vice versa). The method is simple:
draw a random vector, $\gamma$, and evaluate $C(\gamma)$. The correlation
matrix is guaranteed to be positive definite without the need for
additional restrictions.

The new method provides a unified framework for generating random
correlation matrices, including correlation matrices with special
structures. The new method makes it easy to generate random correlation
matrices with wide range of properties: strictly positive elements,
block structures, well-conditioned, in the vicinity of a particular
correlation matrix, and containing elements with similar or heterogeneous
dispersions. In some applications is will be natural for the distribution
on $\mathcal{C}_{n\times n}$ to be invariant to the ordering of the
variables. This would, among other things, imply that the marginal
distributions of the correlation coefficients are identical. This
invariance property is also simple to satisfy with the new method.
Theorem \ref{thm:Permutations} characterizes the class of distributions
for $\gamma$ that leads to random correlation matrices with this
property. Finally, the proposed framework can be used to generate
high dimensional random correlation matrices in a way that is computationally
efficient. 

We have reviewed several existing methods for generating random correlation
matrices. We discussed their advantages and limitations which may
be helpful for selecting the method that is best suited for practical
application. We also identified some peculiar properties of the commonly
used Bendel-Mickey method. 

{\footnotesize{}\bibliographystyle{agsm}
\bibliography{prh}
}{\footnotesize\par}

\appendix

\section{Appendix of Proofs}

\noindent\textbf{Proof of Theorem \ref{thm:Permutations}.} Set $G=\log C(\gamma)$
and consider $\tilde{G}=(PGP)$. Since $G_{ij}=h(\varepsilon_{ij},\xi_{i},\xi_{j},\xi_{0})$
it follows that $\gamma=\mathrm{vecl}(G)$ and $\tilde{\gamma}=\mathrm{vecl}(\tilde{G})$
are identically distributed for any permutation matrix, $P$. Consequently,
$C=C(\gamma)$ and $\tilde{C}=C(\tilde{\gamma})$ are identically
distributed. What remains is to show that $\tilde{C}=PCP^{\prime}$,
which follows from
\[
\tilde{C}=\exp(\tilde{G})=\exp(PGP^{\prime})=P\exp(G)P^{\prime}=PCP^{\prime},
\]
 where we used that $\exp(P\log CP^{\prime})=\exp(PQ\log\Lambda Q^{\prime}P^{\prime})=PQ\exp(\log\Lambda)Q^{\prime}P^{\prime}$
and that $Q^{\prime}P^{\prime}PQ=I$. $\square$

\noindent\textbf{Proof of Theorem \ref{thm:Logistic2Beta}.} Since
$r(z)=\frac{1-e^{-nz}}{1+(n-1)e^{-nz}}$ it follows that $r(z)\in(-\tfrac{1}{n-1},1)$.
Next, we determine the expression for $f_{z}(z)=\left|\tfrac{\partial r(z)}{\partial z}\right|f_{r}(r(z))$,
where 
\[
f_{r}(r)=\frac{1}{B(\alpha,\beta)}\frac{\left(\frac{1}{n-1}+r\right)^{\alpha-1}\left(1-r\right)^{\beta-1}}{\left(\frac{n}{n-1}\right)^{\alpha+\beta-1}}\times1_{\{-\tfrac{1}{n-1}<r<1\}}.
\]
Since $(n-1)e^{-nz}=e^{-\frac{z-\mu}{s}}=e^{-\zeta}$ where $\zeta=\frac{z-\mu}{s}$,
we can write 
\[
r(z)=\frac{1-e^{-nz}}{1+(n-1)e^{-nz}}=\frac{1-\tfrac{1}{n-1}e^{-\zeta}}{1+e^{-\zeta}},
\]
and
\begin{eqnarray*}
\tfrac{1}{n-1}+r(z) & = & \frac{1+e^{-\zeta}+(n-1)(1-\tfrac{1}{n-1}e^{-\zeta})}{(n-1)(1+e^{-\zeta})}=\frac{n}{n-1}\frac{1}{1+e^{-\zeta}},\\
1-r(z) & = & \frac{1+e^{-\zeta}-(1-\tfrac{1}{n-1}e^{-\zeta})}{1+e^{-\zeta}}=\frac{e^{-\zeta}(1+\tfrac{1}{n-1})}{1+e^{-\zeta}}=\frac{n}{n-1}\frac{e^{-\zeta}}{1+e^{-\zeta}}.
\end{eqnarray*}
Since $r(z)\in(-\tfrac{1}{n-1},1)$ is guaranteed, it follows that
\[
f_{r}(r(z))=\frac{1}{B(\alpha,\beta)}\frac{\left(\frac{1}{1+e^{-\zeta}}\right)^{\alpha-1}\left(\frac{e^{-\zeta}}{1+e^{-\zeta}}\right)^{\beta-1}}{\left(\frac{n}{n-1}\right)^{+1}}=\frac{1}{B(\alpha,\beta)}\frac{e^{-\beta\zeta}\left(e^{-\zeta}\right)^{-1}}{\frac{n}{n-1}\left(1+e^{-\zeta}\right)^{\alpha+\beta-2}}.
\]
Next, the derivative is given by
\[
\tfrac{\partial r(z)}{\partial z}=n^{2}\frac{e^{-nz}}{(1+(n-1)e^{-nz})^{2}}=n\frac{n}{n-1}\frac{e^{-\zeta}}{1+e^{-\zeta}},
\]
such that 
\begin{eqnarray*}
f_{z}(z) & = & n\frac{n}{n-1}\frac{e^{-\frac{z-\mu}{s}}}{1+e^{-\frac{z-\mu}{s}}}\frac{1}{B(\alpha,\beta)}\frac{e^{-\beta\frac{z-\mu}{s}}\left(e^{-\frac{z-\mu}{s}}\right)^{-1}}{\frac{n}{n-1}\left(1+e^{-\frac{z-\mu}{s}}\right)^{\alpha+\beta-2}}\\
 & = & \frac{1}{B(\alpha,\beta)}\frac{1}{s}\frac{e^{-\beta\frac{z-\mu}{s}}}{\left(1+e^{-\frac{z-\mu}{s}}\right)^{\alpha+\beta-1}},
\end{eqnarray*}
as stated. This completes the proof. $\square$

\noindent\textbf{Proof of Corollary \ref{Corr:Logistic2Uniform}.}
Follows from Theorem \ref{thm:Logistic2Beta} by setting $\alpha=\beta=1$,
and it can also be verified directly that $f_{z}(z)=\tfrac{n-1}{n}\left|\tfrac{\partial r(z)}{\partial z}\right|=n(n-1)\frac{e^{-nz}}{(1+(n-1)e^{-nz})^{2}}$.
$\square$

{\noindent}\textbf{Proof of Theorem \ref{thm:Block}.} For a given
block structure defined in (\ref{eq:BlockG}) characterized by block
sizes $n_{k}$, $k=1,...,K$, let us introduce a duplication matrix
consisted only of zeroes and ones, $B\in\mathbb{R}^{n\times K}$ ,
such that for any vector $x=(x_{1},...,x_{K})^{\prime}\in\mathbb{R}^{K}$
it holds

\[
B\,x=(\underbrace{x_{1},...,x_{1}}_{n_{1}},\underbrace{x_{2},...,x_{2}}_{n_{2}},...,\underbrace{x_{K},...,x_{K}}_{n_{K}})^{\prime}\in\mathbb{R}^{n},
\]
so the $k$-th element of $x$ appears $n_{k}$ times and $n_{1}+\cdots+n_{K}=n$.
Then, the main diagonal of $G[y]$ takes form of $B\,y$, to be in
agreement with the assumed block structure and is completely determined
by a lower-dimensional vector $y$. Since $G[y]$ has the block structure
(\ref{eq:BlockG}), then $C[y]=\exp G[y]$ has the same block structure
(see \citet{ArchakovHansen:CanonicalBlockMatrix} for details). Therefore,
as in $G[y]$, the entries on the main diagonal of $C[y]$ are identical
within each diagonal block. Introduce vector $q(y)=\Bigl(q_{1}(y),...,q_{K}(y)\Bigl)^{\prime}\in\mathbb{R}^{K}$
which consists of distinct diagonal entries of $C[y]$, where $q_{k}(y)$
is a diagonal element of $k$-th diagonal block from $\exp G[y]$.
Similarly, the main diagonal of $C[y]$ can be expressed as $B\,q(y)$.

Consider a function 

\[
\tilde{g}(y)=\text{diag}(G[y])-\log\text{diag}(\exp G[y])=B\,y-B\,\log q(y)=B\,g(y),
\]
where $\log$ applies element-wise to a vector and $g(y)=y-\log q(y)$.
In \citet{ArchakovHansen:Correlation} it is shown that $\tilde{g}(y)$
is a contraction mapping and converges to a unique diagonal of $G[y]$
such that the main diagonal of $C[y]$ is a vector of ones (and so,
$C[y]$ is a correlation matrix). For our block case, this function
is effectively characterized by a lower-dimensional function $g(y)$,
which is also contraction and converges to the same as $\tilde{g}(y)$
unique fixed-point, $y^{*}$ , since matrix $B$ does not affect the
values of $g(y)$, but only duplicates them. Moreover, from Lemma
2 in \citet{ArchakovHansen:Correlation} it follows that $y_{k}^{*}\leq0$,
for $k=1,...,K$.

Using the canonical representation of block matrices, it is easy to
express a diagonal element of $k$-th diagonal block from $C[y]$,
that is $q_{k}(y)$, as a function of elements from $G[y]$, 

\[
q_{k}(y)=\frac{1}{n_{k}}[\exp\bigl(A+\text{diag}(y)\bigl)]_{kk}+\frac{n_{k}-1}{n_{k}}e^{y_{k}-\gamma_{k,k}},\quad k=1,...,K,
\]
where $[M]_{kk}$ denotes $k$-th diagonal entry of $M$ (see \citet{ArchakovHansen:CanonicalBlockMatrix}
for details). This allows to characterize all elements in the contraction
mapping $g(y)$ from which the algorithm converging to $y^{*}$ follows.

\hfill{}$\square$

\noindent \textbf{Proof of Theorem \ref{thm:PositiveC}.} Consider
$A=G+\alpha I$ where $\alpha=-\min_{i}G_{ii}$, such that $A_{ij}\geq0$
and 
\begin{equation}
[A^{k}]_{ij}=\sum_{h_{1},\dots,h_{k-1}}A_{ih_{1}}A_{h_{1}h_{2}}\cdots A_{h_{k-1}j}\geq0\qquad\text{for all }i,j=1,\ldots,n\text{ and any }k=0,1,\ldots.\label{eq:Ak}
\end{equation}
Since $G$ and $\alpha I$ commute we have $e^{A}=e^{(G+\alpha I)}=e^{G}e^{\alpha I}=e^{G}e^{\alpha},$
such that 
\begin{equation}
C=e^{G}=e^{-\alpha}e^{A}=e^{-\alpha}\sum_{k=0}^{\infty}\frac{1}{k!}A^{k},\label{eq:CalphaA}
\end{equation}
and it follows that $C_{ij}\geq0$ for all $i,j=1,\ldots,n$. 

From (\ref{eq:Ak}) and (\ref{eq:CalphaA}) it follows that $C_{ij}=0\Leftrightarrow[A^{k}]_{ij}=0$
for all $k$$\Leftrightarrow A_{ih_{1}}A_{h_{1}h_{2}}\cdots A_{h_{k-1}j}=0$
for all $h_{1},\ldots,h_{k-1}\in\{1,\ldots,n\}$ and all $k$$\Leftrightarrow$
$A$ is reducible. Since $A$ is reducible if and only if $G$ is
reducible the result follows by contradiction and we can conclude
that $C_{ij}>0$ for all $i,j$ if $\gamma\geq0$ and $G$ is irreducible.
$\square$

\noindent\textbf{Proof of Theorem \ref{thm:Bound}.} From $G=\log C=Q\log\Lambda Q^{\prime}$
it follows that 
\[
G_{ij}=\sum_{k=1}^{n}q_{ik}\log\lambda_{k}q_{kj},
\]
and $G_{ii}\leq0$ for all $i$, because 
\[
G_{ii}=\sum_{k=1}^{n}q_{ik}^{2}\log\lambda_{k}\leq\log\left(\sum_{k=1}^{n}q_{ik}^{2}\lambda_{k}\right)=\log1=0.
\]
Next, with $x_{+}=\tfrac{1}{\sqrt{2}}(e_{i}+e_{j})$ and $x_{-}=\tfrac{1}{\sqrt{2}}(e_{i}-e_{j})$
we find
\[
\log\lambda_{\min}=\min_{\left\Vert x\right\Vert =1}x^{\prime}Gx\leq x_{+}^{\prime}Gx_{+}=\tfrac{1}{2}[G_{ii}+G_{jj}+2G_{ij}]\leq G_{ij},
\]
and similarly $\log\lambda_{\min}\leq-G_{ij}$. By combining the two
inequalities, we have shown the first inequality $\log\lambda_{\min}\leq-\max_{i\neq j}|G_{ij}|$.

Next, order the eigenvalues in descending order, such that $\lambda_{\min}=\lambda_{n}$,
and $(q_{1n},\ldots,,q_{nn})^{\prime}$ is the corresponding eigenvector.
Then
\[
\log\lambda_{\min}=\sum_{ij}q_{in}G_{ij}q_{jn}=\sum_{i\neq j}q_{in}G_{ij}q_{jn}+\sum_{i}q_{in}^{2}G_{ii},
\]
from which it follows that 
\begin{eqnarray*}
|\log\lambda_{\min}| & \leq & \sum_{i\neq j}|q_{in}G_{ij}q_{jn}|+\sum_{i}q_{in}^{2}|G_{ii}|\\
 & \leq & \gamma_{\max}\sum_{i\neq j}|q_{in}q_{jn}|-\sum_{i}q_{in}^{2}G_{ii}\leq\gamma_{\max}(n-1)-\min_{i}G_{ii}.
\end{eqnarray*}
The last term, $-\min_{i}G_{ii}=\max_{i}|G_{ii}|$, is determined
as the fixed-point, $(G_{11},\ldots,G_{nn})=x^{\ast}$, for the contraction
$g(x)=x-\delta(x)$, where $\delta(x)\equiv\log.(\mathrm{diag}(\exp(G[x]))$.
The mapping, $g$, has the Lipschitz constant $\kappa\in(0,1)$, so
that $\kappa=1-1/\Delta$ for some $\Delta>0$. Consider the sequence
$x^{(k+1)}=g(x^{(k)})$. Then $x^{\ast}=\lim_{k\rightarrow\infty}x^{(k)}$
for any $x^{(0)}\in\mathbb{R}^{n}$, moreover we have (from the standard
proof of Banach's fixed-point theorem) that
\[
\left\Vert x^{\ast}-x^{(0)}\right\Vert _{\infty}\leq\frac{1}{1-\kappa}\left\Vert x^{(1)}-x^{(0)}\right\Vert _{\infty},
\]
and we can set $x^{(0)}=0$, such that $x^{(1)}-x^{(0)}=x^{(1)}=-\delta(0)$,
and we have
\[
\max|G_{ii}|=\left\Vert x^{\ast}-0\right\Vert _{\infty}\leq\frac{1}{1-\kappa}\left\Vert \delta(0)\right\Vert _{\infty}\leq\Delta\log(\max_{i}\left|[\exp\{G[0]\}]_{ii}\right|.
\]
Next consider $\tilde{G}_{ij}=\gamma_{\max}$ for all $i\neq j$.
Then 
\[
\max_{i}\left|[\exp\{G[0]\}]_{ii}\right|\leq\max_{i}\left|[\exp\{\tilde{G}[0]\}]_{ii}\right|=\left|[\exp\{\tilde{G}[0]\}]_{11}\right|,
\]
where the inequality follows from the definition of the matrix exponential.
Moreover, 
\[
\tilde{G}[0]=n\gamma_{\max}P-\gamma_{\max}I=n\gamma_{\max}P-\gamma_{\max}(P+P_{\bot})=\gamma_{\max}(n-1)P-\gamma_{\max}P_{\bot},
\]
where $P$ is the $n\times n$ projection matrix with elements $P_{ij}=\frac{1}{n}$
and $P_{\bot}=I-P$. It follows that 
\[
\exp(\tilde{G}[0])=e^{(n-1)\gamma_{\max}}P+e^{-\gamma_{\max}}P_{\bot},
\]
such that the diagonal elements equal, 
\begin{eqnarray*}
[\exp\{\tilde{G}[0]\}]_{11} & = & \frac{e^{(n-1)\gamma_{\max}}+(n-1)e^{-\gamma_{\max}}}{n}\leq e^{(n-1)\gamma_{\max}},
\end{eqnarray*}
and combined we have shown that $\max|G_{ii}|\leq\Delta(n-1)\gamma_{\max}$
and the result follows with $K=\Delta(n-1)$.$\square$%

\section{Expression for a Determinant}

We seek $\psi(C)=\det\tfrac{\partial\gamma}{\partial\varrho}$. Let
$C=Q\Lambda Q^{\prime}$ let $E_{l}\in\mathbb{R}^{d\times n^{2}}$,
$E_{u}\in\mathbb{R}^{d\times n^{2}}$ and $E_{d}\in\mathbb{R}^{n\times n^{2}}$
be the elimination matrices that extract the lower-triangle, upper-triangle,
or diagonal elements of an $n\times n$ matrix, i.e. $\mathrm{vecl}M=E_{l}\mathrm{vec}M$,
$\mathrm{vecl}M^{\prime}=E_{u}\mathrm{vec}M$ and $\mathrm{diag}M=E_{d}\mathrm{vec}M$
for any $M\in\mathbb{R}^{n\times n}$. From \citet[proposition 3]{ArchakovHansen:Correlation}
we have that $\tfrac{\partial\varrho}{\partial\gamma}=E_{l}\Bigl(I-A_{C}E_{d}^{\prime}\Bigl(E_{d}A_{C}E_{d}^{\prime}\Bigl){}^{-1}E_{d}\Bigl)A_{C}(E_{l}+E_{u})^{\prime}$,
were $A_{C}=(Q\otimes Q)\Xi\bigl(Q\otimes Q\bigl)^{\prime}$ and $\Xi$
is the $n^{2}\times n^{2}$ diagonal matrix whose elements are given
by
\begin{equation}
\Xi_{(i-1)n+j,(i-1)n+j}=\xi_{ij}=\begin{cases}
\lambda_{i}, & \text{if}\qquad\lambda_{i}=\lambda_{j},\\
\tfrac{\lambda_{i}-\lambda_{j}}{\log\lambda_{i}-\log\lambda_{j}}, & \text{if}\qquad\lambda_{i}\neq\lambda_{j},
\end{cases}\label{eq:xi_defined-1}
\end{equation}
for $i=1,\ldots,n$ and $j=1,\ldots,n$. Note that $A_{C}$ is symmetric
and positive definite, because $\xi_{ij}>0$ for all $i,j$. Here
we have adapted the expression $\frac{\mathrm{d}\text{vec}\exp X}{\mathrm{d}\text{vec}X}$
in \citet{LintonMcCrorie:1995} to our context where $A_{C}=\frac{\mathrm{d}\text{vec}C}{\mathrm{d}\text{vec}\log C}$.
It follows that 
\[
\psi(C)=\frac{1}{\det\left(E_{l}\Bigl(I-A_{C}E_{d}^{\prime}\Bigl(E_{d}A_{C}E_{d}^{\prime}\Bigl){}^{-1}E_{d}\Bigl)A_{C}(E_{l}+E_{u})^{\prime}\right)}.
\]

\end{document}